\documentclass[aip,sd,amsmath,amssymb,reprint]{revtex4-1}
\usepackage{graphicx}
\usepackage{dcolumn}
\usepackage{bm}
\usepackage{{subfig}}
\usepackage{color}

\begin{document}

\preprint{}

\title{Charge regulation in ionic solutions: thermal fluctuations and Kirkwood-Schumaker interactions} 
\author{Nata\v sa Ad\v zi\' c}
\email[]{natasa.adzic@ijs.si}

\affiliation{Department of Theoretical Physics, J. Stefan Institute, 1000 Ljubljana, Slovenia.}

\author{Rudolf Podgornik}
\affiliation{Department of Theoretical Physics, J. Stefan Institute, and
Department of Physics, Faculty of Mathematics and Physics, University of Ljubljana, 1000 Ljubljana, Slovenia.}

\date{\today}

\begin{abstract}
We study the behavior of two macroions with dissociable charge groups, regulated by local variables such as pH and electrostatic potential, immersed in a mono-valent salt solution, considering cases where the net charge can either change sign or remain of the same sign depending on these local parameters. The charge regulation, in both cases, is described with the proper free energy function for each of the macroions, while the coupling between the charges is evaluated on the approximate Debye-H\"uckel level. The charge correlation functions and the ensuing charge fluctuation forces  are calculated analytically and numerically. Strong attraction between like-charged macroions is found close to the point of zero charge, specifically due to {\sl asymmetric, anticorrelated charge fluctuations} of the macroion charges. The general theory is then implemented for a system of two protein-like macroions, generalizing the form and magnitude of the Kirkwood-Schumaker interaction.
\end{abstract}


\maketitle

\section{Introduction}\label{sec:1}

From the point of view of electrostatic interactions, proteins, as ampholytes,  are challenging objects since they carry a non-constant charge, dependent on dissociation of chargeable molecular moieties such as N- and C-terminals as well as the (de)protonation of amino acid side groups \cite{Piazza,Simonson,Leckband}. Consequently, their behavior can not be analyzed with the assumption of a constant charge \cite{BoJ}, otherwise applicable for many (bio)colloidal systems \cite{RudiandCo,Lebovka}, since it misses the crucial contribution of charge regulation and charge fluctuations to the interactions between macroions \cite{Borkovec-review}. In fact, it has been known for some time that extremely long-ranged attractive interactions occur between proteins in an aqueous solution close to the point of zero charge (PZC), as first elucidated by Kirkwood and Shumaker \cite{KS1, KS2}. The approximate form of the Kirkwood-Shumaker (KS) interaction is fundamentally different from the van der Waals (vdW) interactions \cite{Pit}, that stem only from dipolar fluctuations and act between electro-neutral bodies, since it is a consequence of the {\sl monopolar charge fluctuations} and does not exist for macroions with a strictly fixed charge distribution.  KS interaction therefore pertains only to systems with flexible charge equilibrium that posses a non-zero capacitance, where the net charge is not a constant but depends on the underlying dissociation processes \cite{Borkovec-more}. This furthermore implies that the effective charge on the macroion, e.g. the protein surface, is regulated and responds to the local solution conditions: $pH$, electrostatic potential, salt concentration, spatial dielectric constant profile and the presence of other vicinal charged groups \cite{Lund}. While the effects of charge regulation were analyzed on various levels in the mean-field approximation \cite{Chan,Me-old,ctab,Carnie,Szleifer,Boon,Borkovec-more,Netz-CR}, the fluctuation effects have not received a proportional attention.

Recently, the KS theory experienced renewed interest when it was shown, using detailed Monte-Carlo simulations \cite{BoJ}, that indeed there exists an interaction between proteins which has the same salient features as the original approximate form of the KS interaction. An important step further was achieved by consistently including the charge regulation free energy \cite{Borkovec-more}, derivable from the Parsegian-Ninham model \cite{Pars-CR}, into the theoretical framework that allowed to derive analytically and exactly the interaction free energy on the Gaussian fluctuation level \cite{ja}, leading to an exact form of the KS interaction for the 3-dimensional system with planar geometry. The full exact solutions for charge regulation interaction, beyond the Gaussian fluctuation {\sl Ansatz}, have been found also in the case of a family of 1-dimensional models solvable by the transfer matrix formalism \cite{Maggs}. 

The aim of this paper is to present a theory of fluctuation interaction in the asymptotic regime of large separations for two small spherical macroions subject to charge regulation. The problem is formulated in the way that allows for decoupling of the charge regulation part and the interaction part, of which the former can be treated exactly while the latter can be dealt with on the Debye-H\" uckel (DH) level. This allows us to derive a closed form expression for the total interaction and compare it with various approximate forms, including the original KS expression. Furthermore we are able to go beyond the KS approximation and derive realistic pH and ionic strength dependent  interactions between protein macroions with known amino acid composition.		

The paper is arranged as follows: in Section \ref{sec:2}, we introduce a model consisting of two spherical macroions immersed in a mono-valent salt solution, with charge regulated surface charges described with an appropriate free energy term. The theory of electrostatic interactions for such a system is derived using the field-theoretical approach, described in Appendix \ref{sec:app}. Three different cases of charge regulation are considered, Section \ref{sec:3}: a fully symmetric system, consisting of two identical macroions with both charges spanning the interval $[-e, e]$, a semi-symmetric system, composed of two identical macroions, with charges spanning the asymmetric interval $[-e, (\alpha -1)e]$, ($\alpha >1$), and a completely asymmetric system composed of one negative and one positive macroion, with charges $[-e, 0]$ and $[0, e]$, respectively. For all three cases we calculate the average charge, the charge-charge cross-correlation function, the charge-charge auto-correlation function, as well as the total interaction potential obtained numerically using the exact evaluation of the full partition function as well as {\sl via} two symplifying and illuminating approximation methods, Section \ref{sec:4}:  the saddle-point method and the Gaussian approximation method, both giving an analytical closed form for the full charge regulation interaction, including the thermal fluctuations. In Section \ref{sec:5}, we show how this theory can be generalized to be applicable to a system of protein-like macroions with specific amino acid composition. Finally, in section \ref{sec:6} we present our conclusions and comment on the connection with experiments/simulations.

\section{Model}\label{sec:2}

\begin{figure}
\centering{\includegraphics[width=0.4\textwidth]{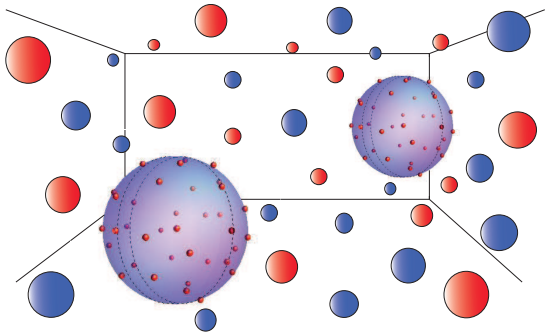}}
\caption{Shematic representation of the model: two charge-regulated ions immersed in 1:1 salt solution.}
\label{fig:fig1}
\end{figure}

We consider a model system composed of two charged spherical macroions in a 1:1 salt solution, Fig \ref{fig:fig1}. The charge of the macroions is not constant, but is described by a dissociation surface free energy cost corresponding to the Parsegian-Ninham charge regulation model, as discussed in \cite{ja}, of the general lattice gas form
\begin{equation}
 f_0(\varphi ({\bf r}))=i\sigma_0\varphi ({\bf r}) -\alpha k_BT\frac{\sigma_0}{e_0}\ln{(1+b e^{i\beta e_0\varphi ({\bf r})})},\label{eq:2}
\end{equation}
where $\alpha$ quantifies the number of dissociation sites and $\ln{b} = -\ln{10}(pH-pK) = { \beta \mu _S}$, where $pK$ is the dissociation constant and $\mu _S$ is the free energy of charge dissociation. Here $\varphi ({\bf r})$ is the local fluctuating potential that needs to be integrated over to get the final partition function. The mean-field Poisson-Boltzmann (PB) approximation is obtained by identifying $\varphi ({\bf r}) \longrightarrow i \phi = i \phi_{PB}$ \cite{ja}. The total dissociation free energy for a spherical macroion of a radius $a_0$, sufficiently small so that one can assume that the electrostatic potential is uniform over its surface, $\varphi (\vert {\bf r}\vert = a_0)=\varphi $,  and can be written in the form
\begin{eqnarray}
f(\varphi ) &=& \oint_{S} f_0(\varphi ({\bf r})) d^2{\bf r}  \longrightarrow \nonumber\\
& & \longrightarrow i N e_0\varphi -\alpha k_BTN\ln{(1+b e^{i\beta e_0\varphi})},\label{eq:2}
\end{eqnarray}
where N is the number of absorption sites satisfying $\int dS \sigma _0=Ne_0$, and $\alpha > 1$ is a coefficient of asymmetry, determining the width of the interval spanned by the particle's effective charge $e(\phi)$ as a function of the  mean-field potential on its surface $\phi = \phi (a_0)$:
\begin{eqnarray}
&& e(\phi = \phi (a_0)) =\frac{\partial f(\phi)}{\partial \phi}=\nonumber\\
&& e_0 N\left((\frac{\alpha }{2}-1)- \frac{\alpha }{2}\tanh{[-\frac{1}{2}(\ln{b}-\beta e_0\phi )]} \right).
\label{vaeht}
\end{eqnarray}
The effective charge of the macroion can thus fluctuate in the interval $-Ne_0< e <(\alpha -1)Ne_0$, $\alpha >1$. When $\alpha =2$ the charge interval is by definition symmetric $[-Ne_0, Ne_0]$. All of the expressions for the charge regulation referred to above are just variants of the surface lattice gas free energies \cite{ja} with a variable number of dissociation sites that describe the dissociation of the charge moieties on the surface of the macroions. In addition we have taken the limit of small macroions, implying that the surface potential on the macroions is a constant, $f(\varphi) = \oint_{\vert{\bf r}\vert = a_0} f_0(\varphi ({\bf r})) d^2{\bf r}  $. 

Assuming that the fluctuating electrostatic potential of one macroion is $\phi_1 (a) = \varphi _1$ and of the other one is $\phi_2 (a) = \varphi _2$, located at $\ \vec{r}_1$ and  $\ \vec{r}_2$ respectively, the partition function of the system can be derived in the field-theoretic form, see Appendix \ref{sec:app}:
\begin{equation}
{\cal{Z}}=
\int\!\!\int d\varphi _1e^{-\beta f(\varphi _1)} G(\varphi_1,\varphi_2) e^{-\beta f{(\varphi _2)}}d\varphi _2,\label{eq:path}
\end{equation}
where the partition function has already been normalized by dividing with the bulk system partition function \cite{li}, obtained for $f (\varphi )=0$. $G(\varphi _1, \varphi _2)$ is the propagator of the field, defined with the values of the potential  $\varphi_1$ and $\varphi_2$  at the location of the first and the second particle respectively, derived in Appendix \ref{sec:app}:
\begin{eqnarray}
G(\varphi_1,\varphi_2)=e^{-\frac{\beta }{2}\left(\varphi_1, \varphi_2\right)\Big(
\begin{matrix}
G(\vec{r}_1,\vec{r}_1) &  G(\vec{r}_1,\vec{r}_2)\\
G(\vec{r}_1,\vec{r}_2) &  G(\vec{r}_2,\vec{r}_2)
\end{matrix}\Big)^{-1}
\left(\begin{array}{c} \varphi_1 \\ \varphi_2 \end{array}\right)},\nonumber\\
~
\end{eqnarray}
where the matrix of Green's functions for the bulk composed of a 1:1 electrolyte in the DH approximation is given as:
\begin{equation}
\Bigg(\begin{matrix}
G(\vec{r}_1,\vec{r}_1) &  G(\vec{r}_1,\vec{r}_2)\\
G(\vec{r}_1,\vec{r}_2) &  G(\vec{r}_2,\vec{r}_2)
\end{matrix}\Bigg)=\Bigg(\begin{matrix}
\frac{1}{4\pi\epsilon\epsilon_0}\frac{e^{-\kappa a}}{a} & \frac{1}{4\pi\epsilon\epsilon_0} \frac{e^{-\kappa R}}{R}\\
\frac{1}{4\pi\epsilon\epsilon_0}\frac{e^{-\kappa R}}{R} & \frac{1}{4\pi\epsilon\epsilon_0} \frac{e^{-\kappa a}}{a}
\end{matrix}\Bigg),
\end{equation}
Here we assumed that the two macroions can not come closer then $a=2a_0$. Variations on the above form are possible that would contain the factor $\frac{e^{-\kappa (R - a)}}{R (1 + \kappa a)}$ for the separation dependence of $G(\vec{r},\vec{r})$. We will comment on the detailed choice of the form for the DH interaction later.

The charge regulation energy term $\ e^{-\beta f(\varphi ) }$ can now be expanded as a binomial \cite{Maggs}:
\begin{eqnarray}
e^{-\beta f(\varphi ) }&=&e^{-i \beta N e_0\varphi }(1+b e^{i\beta e_0\varphi})^{\alpha N}=\nonumber\\
&&\sum_{n=0}^{\alpha N}\left(\begin{array}{c} \alpha N \\ n \end{array}\right)b^{n} e^{-i \beta N e_0\varphi } e^{i\beta e_0n\varphi}.
\end{eqnarray}
Integral (\ref{eq:path}) then becomes:
\begin{eqnarray}
&&{\cal{Z}}=\frac{1}{{\cal{Z}}_0}\int \int d\varphi _1d\varphi _2\sum_{n}^{\alpha N}\sum_{n'}^{\alpha N} a_n a_{n'} e^{-i\beta e_0(N-n)\varphi _1}\times\nonumber\\
&& e^{-\frac{\beta }{2}\left(\varphi_1, \varphi_2\right)\Big(
\begin{matrix}
G(\vec{r}_1,\vec{r}_1) &  G(\vec{r}_1,\vec{r}_2)\\
G(\vec{r}_1,\vec{r}_2) &  G(\vec{r}_2,\vec{r}_2)
\end{matrix}\Big)^{-1}\left(\begin{array}{c} \varphi_1 \\ \varphi_2 \end{array}\right)} e^{-i\beta e_0(N-n')\varphi _2}\nonumber\\
~\label{eq:dsjkgjbv}
\end{eqnarray}
where $\ a_n(\alpha)=\left(\begin{array}{c} \alpha N \\ n \end{array}\right)b^{n}$ for any $\alpha$.

Introducing the dimensionless variables $\ \tilde{R}=\kappa R$, $\tilde{a}=\kappa a$, one can rewrite the partition function for two equal macroions with both charges allowed to vary in the interval $[-Ne_0, Ne_0]$ in the form:
\begin{eqnarray}
{\cal{Z}}=\sum_{n}^{2 N} \sum_{n'}^{2 N}   a_n(2) a_{n'}(2) e^{-\beta {\cal F}_{N,N} (n, n', \tilde{R})}, 
\label{eq:partfunction}
\end{eqnarray}
where we introduced ${\cal F}_{N,N} (n, n', \tilde{R})$ as:
\begin{eqnarray}
&& {\cal F}_{N,N} (n, n', \tilde{R})=\frac{e_0^2\kappa }{8\pi \epsilon\epsilon_0}\times \nonumber\\
 &&\left(\frac{e^{-\tilde{a}}}{\tilde{a}}[(N-n)^2+(N-n')^2]+2\frac{e^{-\tilde{R}}}{\tilde{R}}(N-n)(N-n')\right).\nonumber\\
 ~
\end{eqnarray}
Clearly we have incorporated exactly the charge regulation free energy for each of the macroions, while the electrostatic coupling between the two macroions is included approximately {\sl via} the DH propagator. The configuration of this particular example is symmetric, as the two macroions are identical and are descibed by the same charge regulation free energy. The asymmetric configuration, corresponding to unequal charge regulation free energies for the two macroions, is  addressed next.  

In order to describe two equal macroions with a regulated charge in the interval $-Ne_0< e <0$ we take as a model expression  Eq. \ref{eq:2} with $\alpha = 1$, {\sl i.e.}, 
\begin{equation}
f(\varphi )=i M e_0\varphi - \alpha k_BTN \ln{(1+b e^{i\beta e_0\varphi})},
\end{equation}
where $M=N$ and with the partition function
\begin{eqnarray}
{\cal{Z}}=\sum_{n}^{N} \sum_{n'}^{N}   a_n(1) a_{n'}(1) e^{-\beta {\cal F}_{N,N} (n, n', \tilde{R})}.
\label{eq:partfunction}
\end{eqnarray}
Furthermore, charge regulation in the interval $0< e < Ne_0$ is described with 
\begin{equation}
f(\varphi )= -k_BTN\ln{(1+b e^{i\beta e_0\varphi})},
\label{fhjwk}
\end{equation}
corresponding to the protonisation of neutral state ($M=0$), with the partition function for two equal macroions obtained in the form:
\begin{equation}
{\cal{Z}}=\sum_{n=0}^{ N} \sum_{n'=0}^{ N} a_n(1) a_{n'}(1) e^{-{\cal F}_{0,0} (n, n', \tilde{R})}.\\
~\label{eq:apartf}
\end{equation}
Finally, for an asymmetric case where the two macroioins are different, one with charge in the allowed interval $[0, Ne_0]$ and the other one spanning the interval $[-Ne_0, 0]$, the partition function is obviously obtained in the form 
\begin{equation}
{\cal{Z}}=\sum_{n=0}^{ N} \sum_{n'=0}^{ N} a_n(1) a_{n'}(1) e^{-{\cal F}_{N,0} (n, n', \tilde{R})},\\
~\label{eq:apartf}
\end{equation}
These results for the partition function derived above can be written succinctly in a single formula as:
 \begin{equation}
{\cal{Z}}=\sum_{n}^{\alpha N} \sum_{n'}^{\alpha N} a_n(\alpha) a_{n'}(\alpha) ~e^{-\beta  {\cal F}_{N,M}(n, n', \tilde{R})},\\
~\label{eq:aspartfunction}
\end{equation}
where one can distinguish three different cases: 
\begin{itemize}
\item a) $M=N$, $\alpha =2$ - full symmetric system (the macroions are identical, both with charge spanning the symmetric interval $[-Ne_0, Ne_0]$);
\item b) $M=N$, $\alpha >2$ - semi-symmetric system (the macroions are identical, both with charge spanning the asymmetric interval $[-Ne_0, \alpha Ne_0]$);
\item c) $N \neq 0$, $M=0$, $\alpha =1$ - asymmetric system (one particle is positive, with charge fluctuating $[0, Ne_0]$, the other negative with charge spanning the interval $[-Ne_0, 0]$).
\end{itemize}
The partition function Eq. \ref{eq:aspartfunction} can be evaluated exactly only numerically, which is what we will do, but also provide two approximate methods that yield explicit analytical approximations to the exact evaluation.

\section{Symmetric-asymmetric charges on proteins}\label{sec:3}

\begin{figure*}[t!]
\begin{center}
\centering{\subfloat[]{\includegraphics[width=0.34\textwidth]{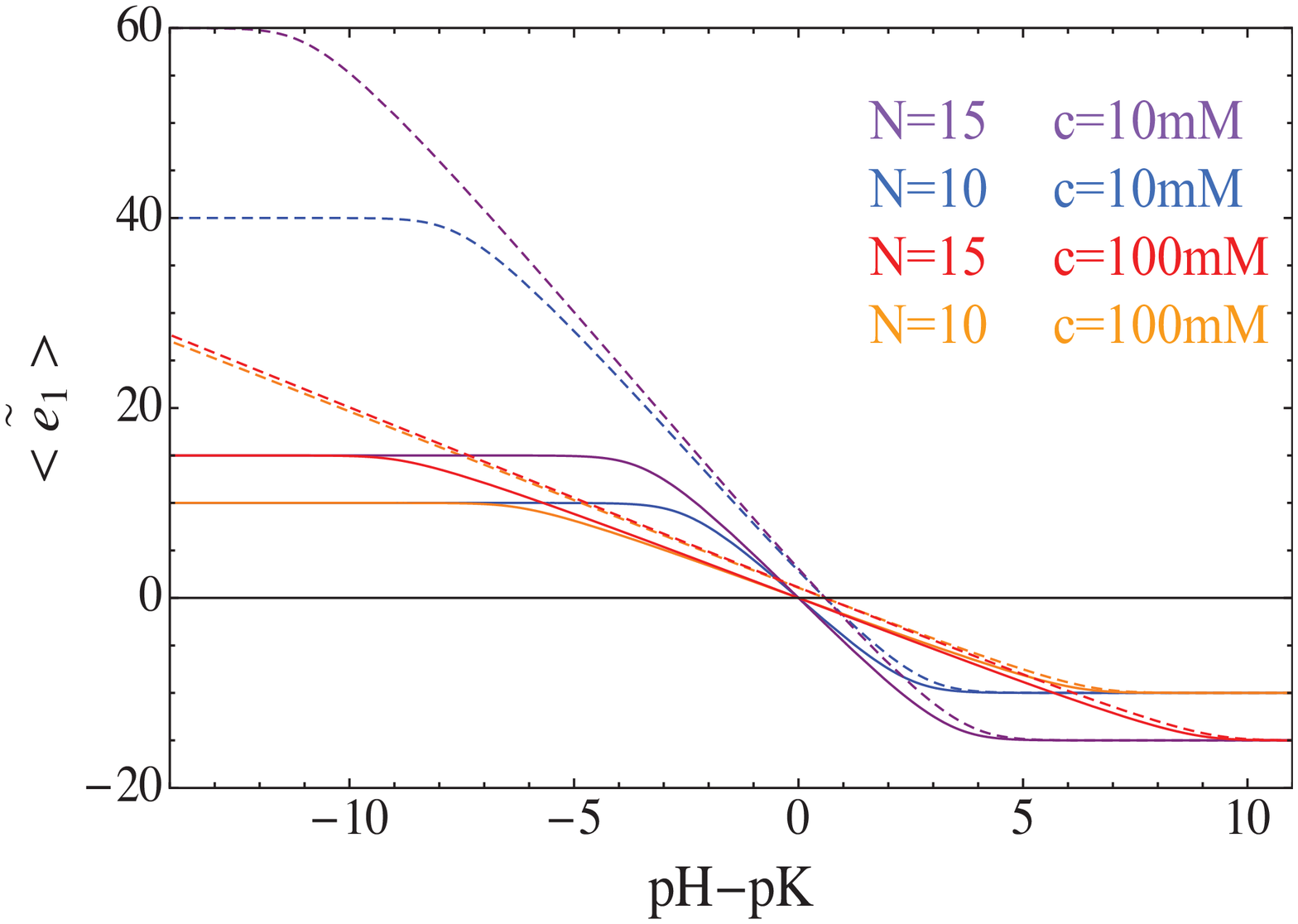}}\subfloat[]{\includegraphics[width=0.34\textwidth]{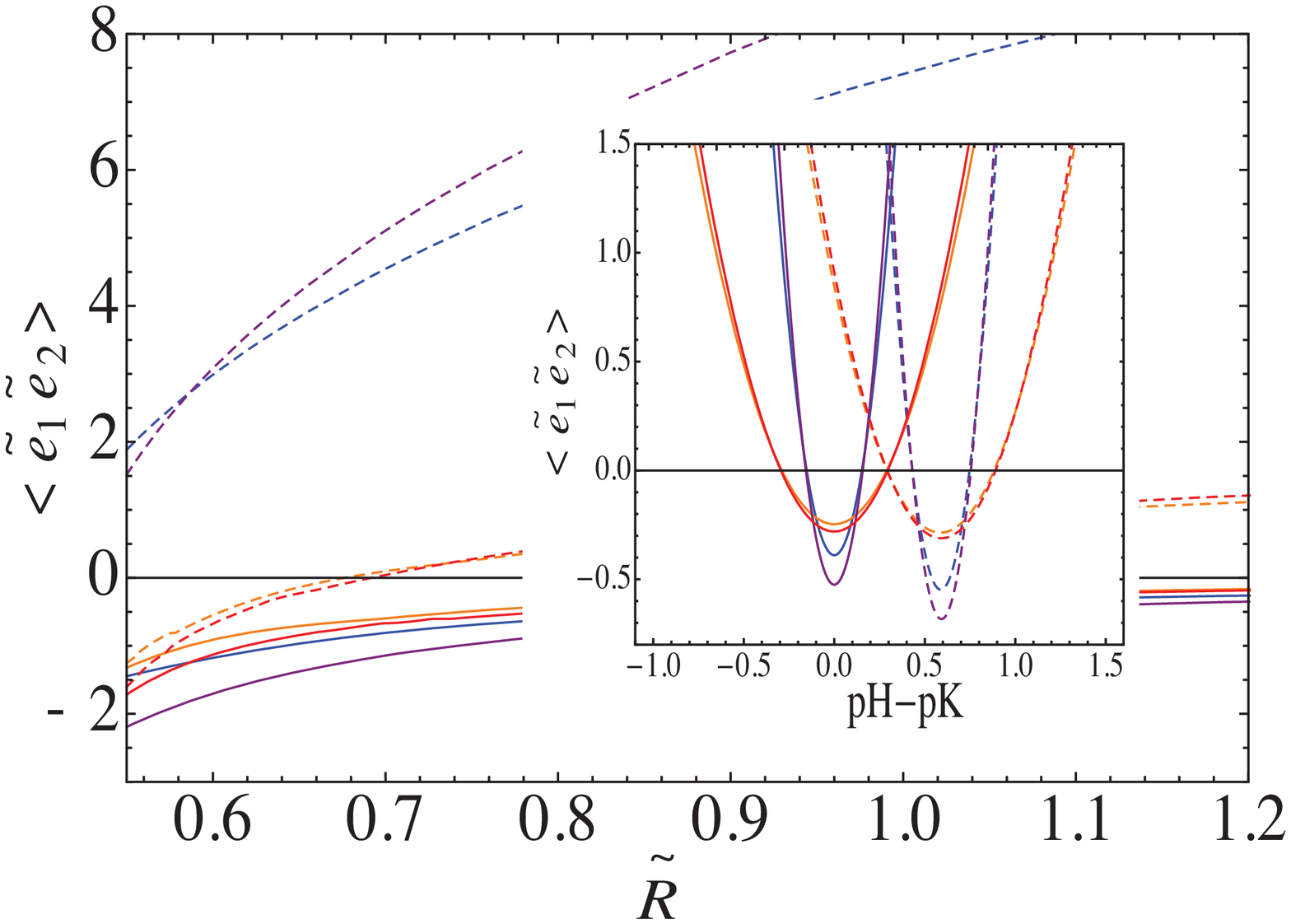}}\subfloat[]{\includegraphics[width=0.34\textwidth]{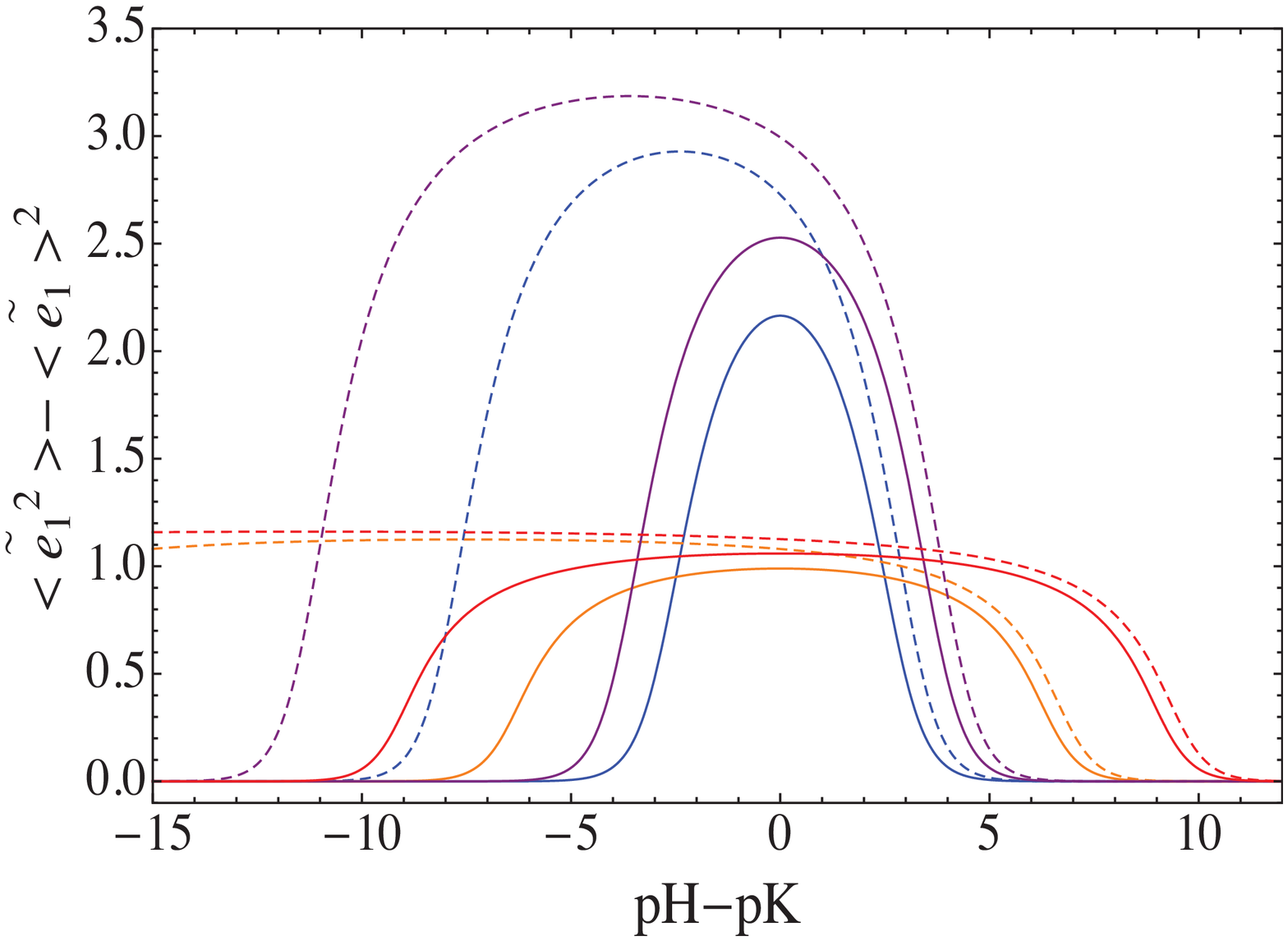}}}
\caption{Symmetric system: (a) The average charge of macroions;  (b) charge cross-correlation function; (c) auto-correlation function. All averages are obtained by exact evaluation of the partition function. Solid lines correspond to a fully-symmetric system ($\alpha =2$), while dashed lines represent the semi-symmetric case which takes asymmetry coefficient to be $\alpha =5$.  Each color corresponds to a  choice of parameters (number of adsorption sites $N$ and salt concentration $c$) as described in (a). The dimensionless diameter of the macroions is set to be $\tilde{a}=0.5$ and separation between them $\tilde{R}=1$. The $\tilde{R}$ dependence is plotted at the PZC, $pH-pK=0$.}
\label{fig:fig2}
\end{center}
\end{figure*}

\begin{figure}[t!]
\centering{\includegraphics[width=0.46\textwidth]{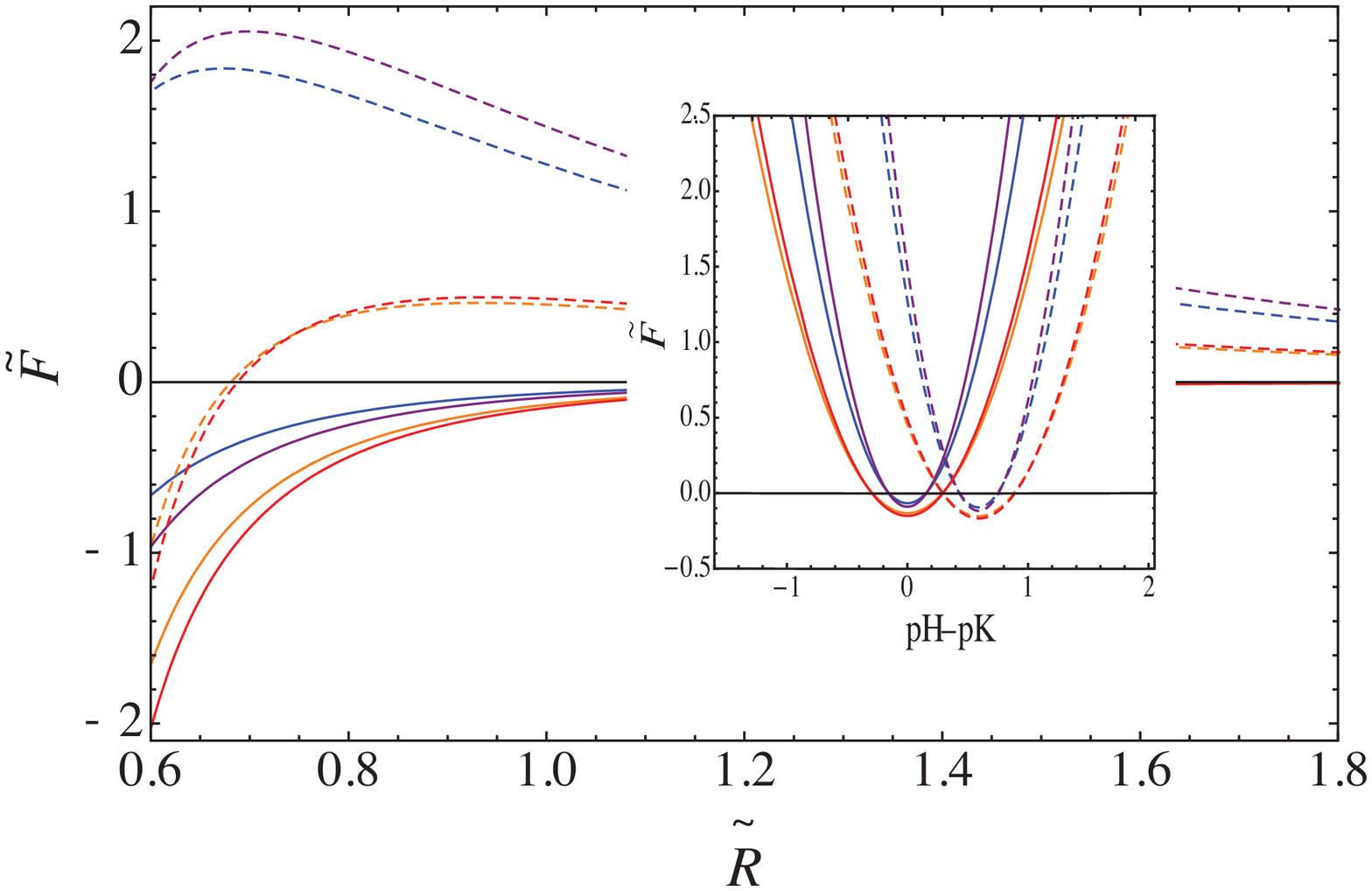}}
\caption{The interaction force for fully-symmetric system (solid lines) and semi-symmetric system (dashed lines). All averages are obtained by exact evaluation of the partition function.  Each color corresponds to a  choice of parameters (number of adsorption sites $N$ and salt concentration $c$) as described in (a). The $\tilde{R}$ dependence is plotted at the PZC, $pH-pK=0$, while the $pH-pK$ dependence is plotted setting $\tilde{R}=1$. The dimensionless diameter of the macroions is taken to be $\tilde{a}=0.5$. }
\label{fig:fig22}
\end{figure}

\begin{figure*}
\centering{\subfloat[]{\includegraphics[width=0.36\textwidth]{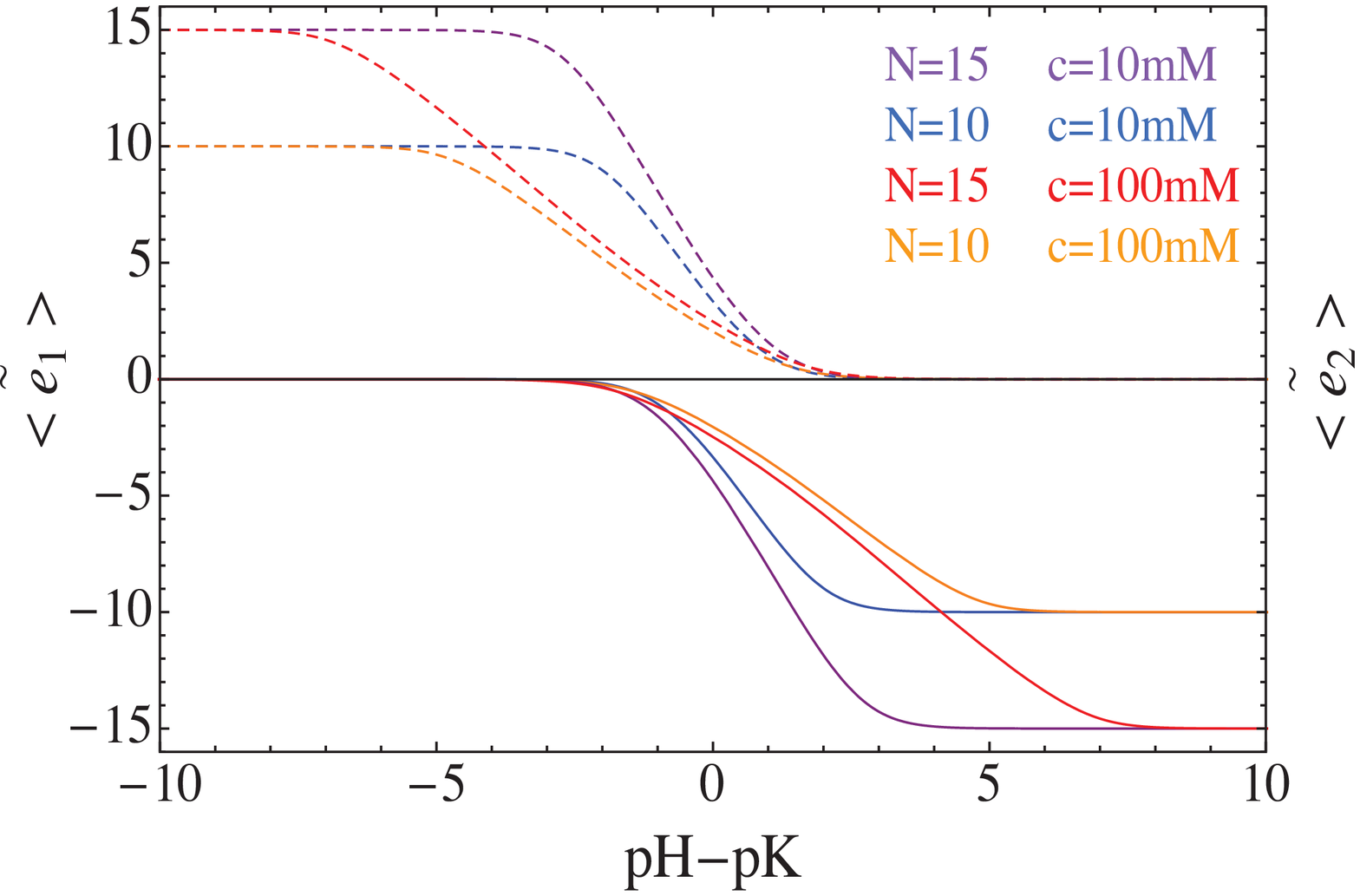}}\subfloat[]{\includegraphics[width=0.34\textwidth]{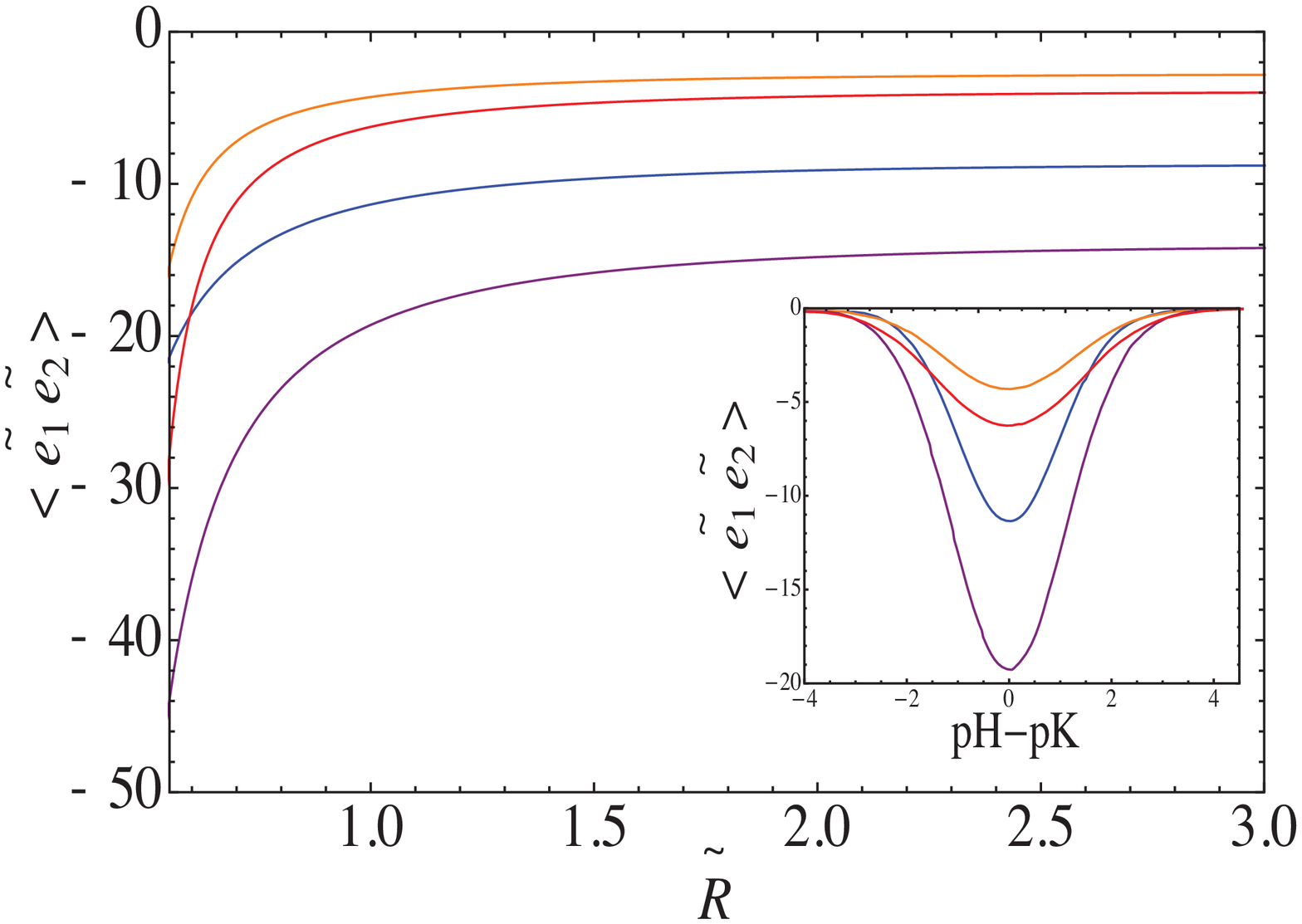}}
\subfloat[]{\includegraphics[width=0.34\textwidth]{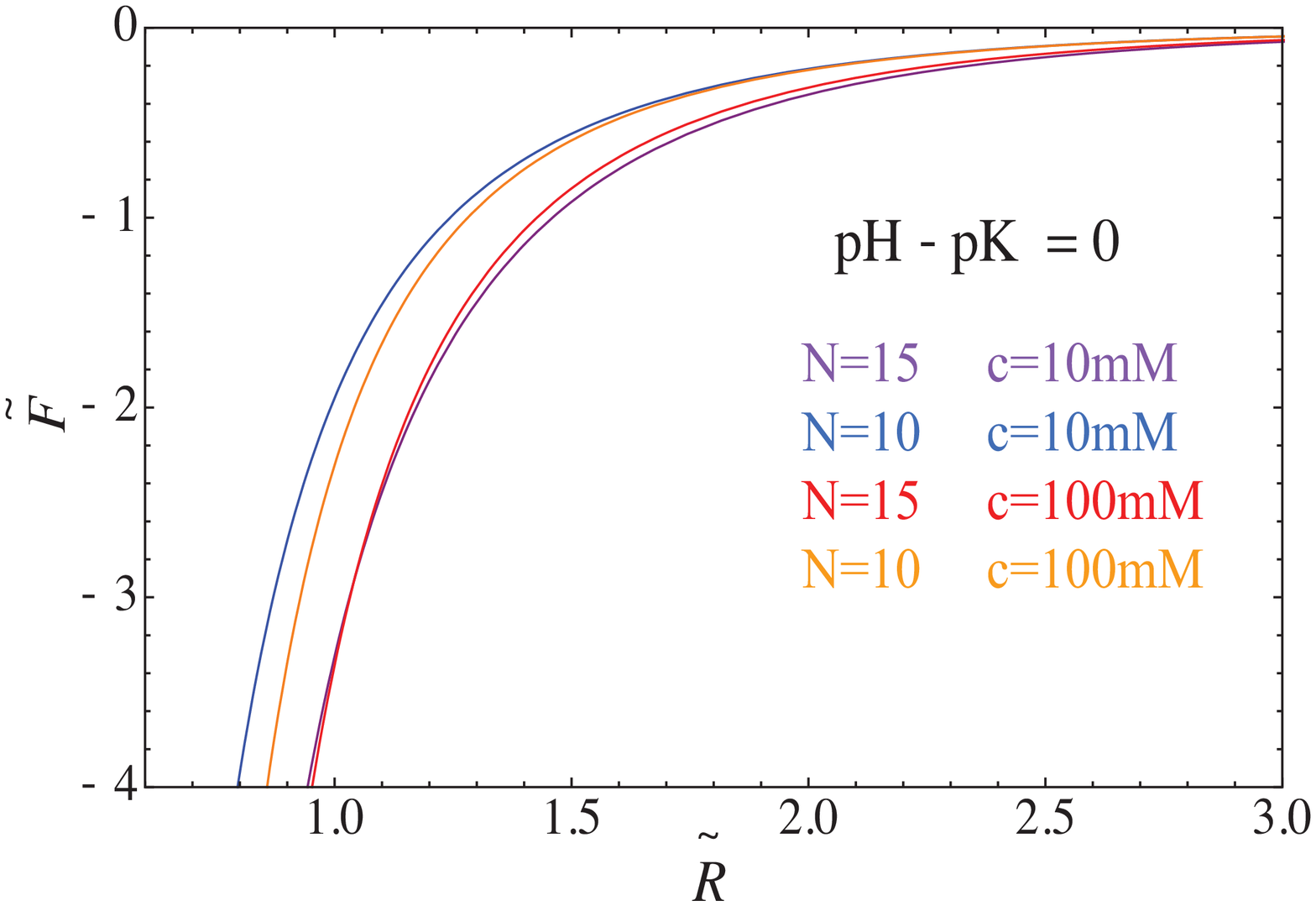}}}
\caption{Asymmetric system: (a) The average charge of one particle (solid lines) and the other (dashed lines); (b) charge cross-correlation function; (c) the interaction force. All averages are obtained using the exact evaluation of full partition function.  Each color corresponds to a choice of parameters (number of adsorption sites $N$ and salt concentration $c$) as described in (a). The dimensionless  diameter of the macroions is set to be $\tilde{a}=0.5$ and the separation between them $\tilde{R}=1$. The $\tilde{R}$ dependence is plotted at point determined with $pH-pK=0$.}
\label{fig:fig3}
\end{figure*}

\begin{figure}[t!]
\subfloat[]{\includegraphics[width=0.48\textwidth]{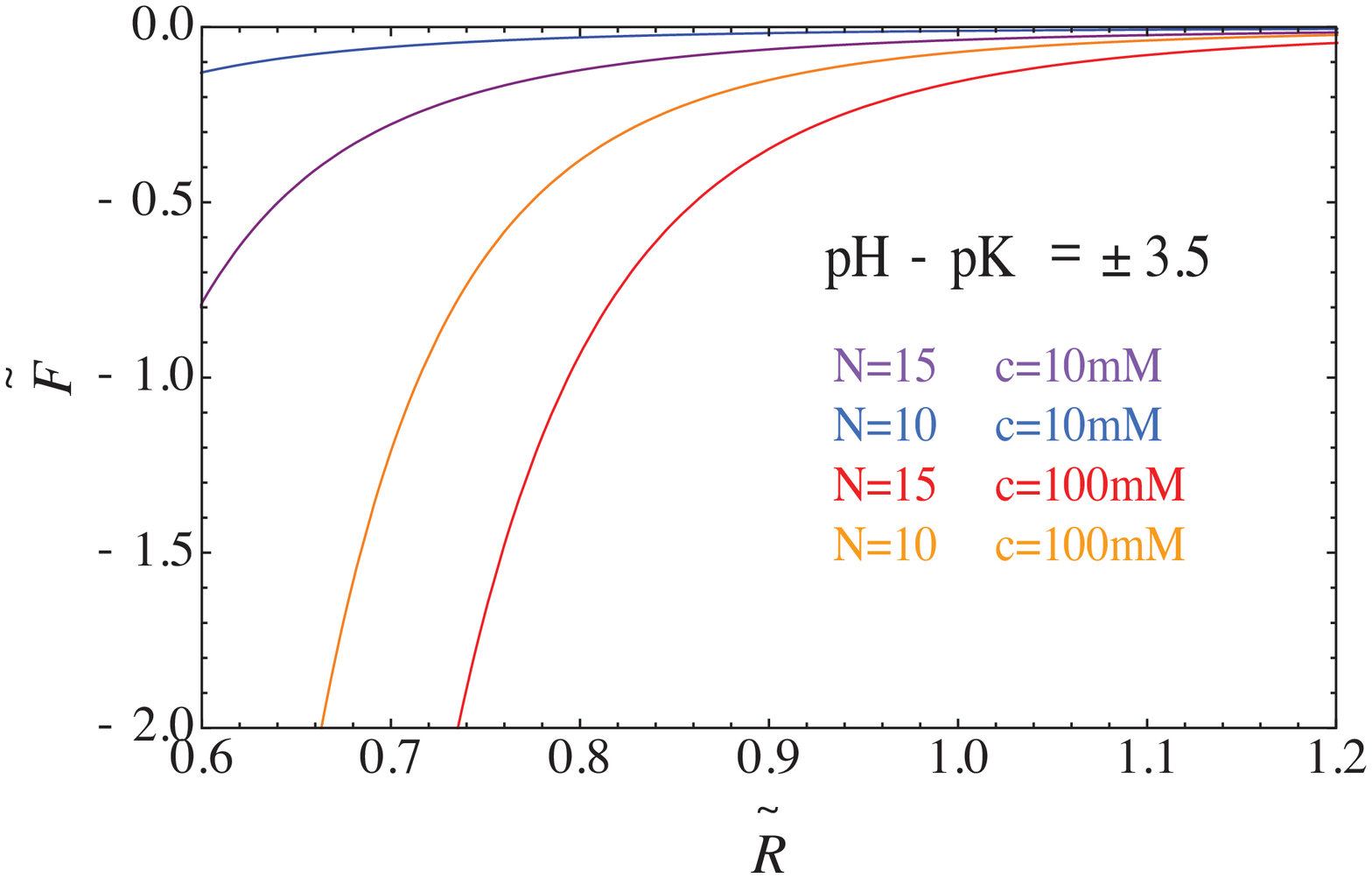}}\\
\subfloat[]{\includegraphics[width=0.5\textwidth]{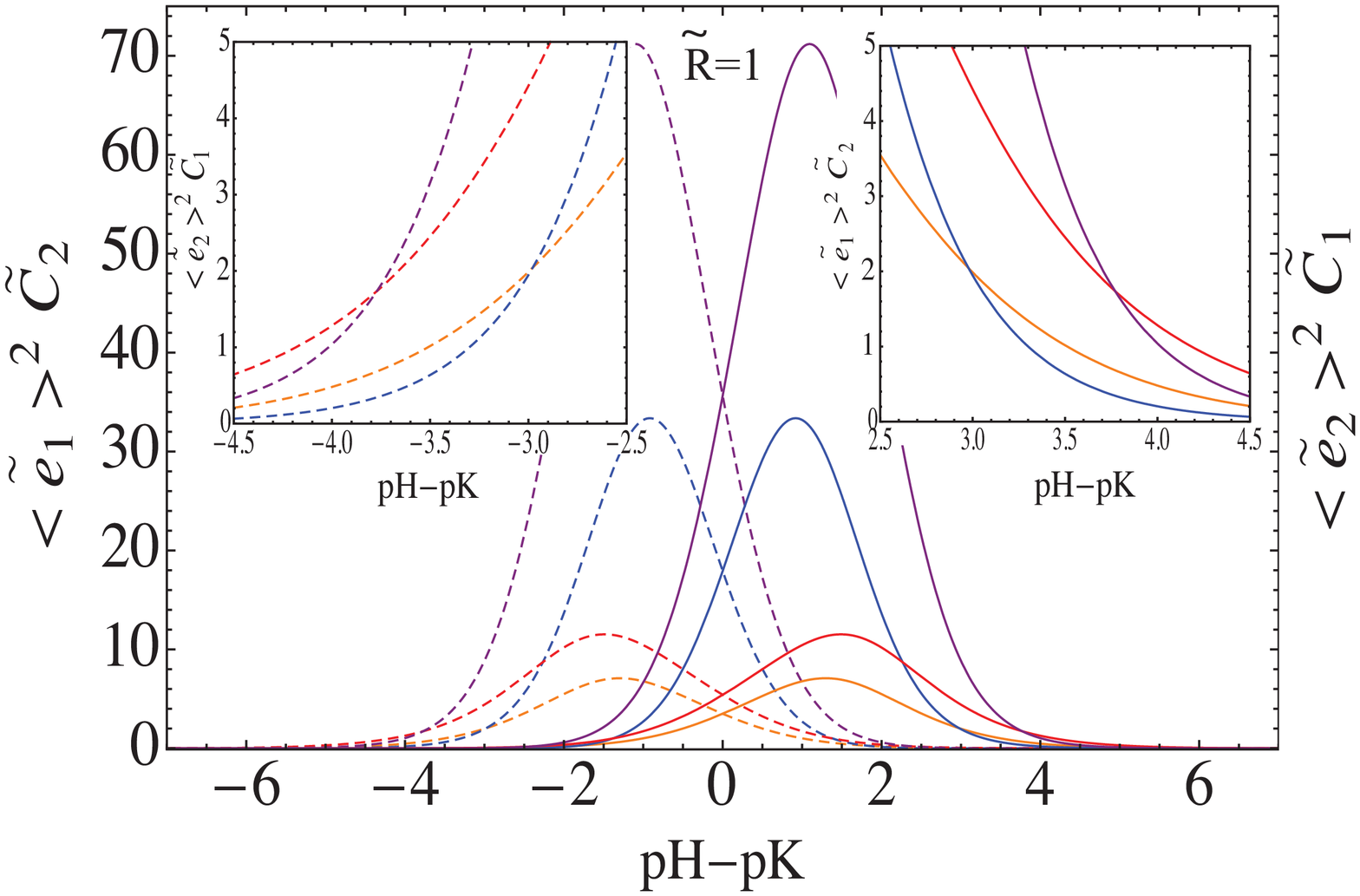}}
\caption{Asymmetric system:  (a) the interaction force plotted at $pH-pK=3.5$; (b)  $<\tilde{e}_1>^2 (<\tilde{e}_2 - <\tilde{e}_2>>^2)$ (solid lines) and $<\tilde{e}_2>^2 (<\tilde{e}_1 - <\tilde{e}_1>>^2)$ (dashed lines).  All averages are obtained using the exact evaluation of full partition function.  Each color corresponds to a choice of parameters (number of adsorption sites $N$ and salt concentration $c$) as described in (a). The dimensionless  diameter of the macroions is set to be $\tilde{a}=0.5$. }
\label{fig:fig33}
\end{figure}

With this we proceed to calculate the average value of the charge of the macroions $\ <e_{1, 2}>$,  charge cross correlation  $\ <e_1e_2>$ and  auto-correlation function $<e_1 - <e_1>>^2$ for all three systems. The thermodynamic averages can be written as
\begin{eqnarray}
<\dots > = \frac{1}{{\cal Z}}\sum_{n, n'}^{\alpha N}a_n(\alpha) a_{n'}(\alpha) \dots e^{-\beta {\cal F}_{N,M}(n, n', \tilde{R})}. ~~~~
\end{eqnarray}
In this way we can write e.g. the dimensionless average charge of the particle, $<\tilde{e}_1>=<e_1>/e_0$ as:
\begin{eqnarray}
<\tilde{e}_1>=<(n-M)>.
\end{eqnarray}
In a similar way, other averages are calculated exactly from the full partition function and are plotted as functions of $\tilde R$ and $pH-pK$, for different values of the number of absorption sites $N$ and salt concentration $c$, keeping fixed the diameter of the macroions $\tilde{a}$, see  Fig. \ref{fig:fig2} and Fig. \ref{fig:fig3}. 

In a fully symmetric system, Fig \ref{fig:fig2} (solid lines), the average charge is allowed to vary in a symmetric interval, reaching the point of zero charge (PZC) for $pH=pK$.  Away from PZC, the average charge changes almost linearly until it reaches saturation and stays constant for any value of $pH-pK$, Fig \ref{fig:fig2}(a). The charge cross correlation function, being negative close to the PZC, indicates that even in the fully symmetric system the macroion charges prefer to fluctuate asymmetrically: charge fluctuation on one macroion being accompanied with a fluctuation of the opposite sign on the other macroion, Fig \ref{fig:fig2}(b). This is a robust property of the system, fully discernable also in the 1-dimensional exact solutions \cite{Maggs}. Considering the charge cross correlation function as a function of distance between macroions,  plotted for fixed $pH-pK$, one can observe that at the PZC, fluctuation asymmetry effect decreases as separation increases, and it is the strongest for smaller values of salt concentration, while close to PZC, the asymmetry appears in regime of larger salt concentration and smaller separations.  The charge auto-correlation function is positive with the maximum centered at the PZC, being bigger for smaller salt concentration, Fig \ref{fig:fig2}(c). 

Finally, the interaction force is calculated as $$\tilde F (\tilde R)= - \frac{d}{d \tilde R}(-\ln{{\cal Z}(\tilde R)}),$$ and it is shown in Fig \ref{fig:fig22}. Two identical macroions repel for most values of the parameters, but show a net attraction in the vicinity of the PZC. This attraction is of purely fluctuational origin, stemming from the asymmetric charge cross-correlation. At the same value of dimensionless separation, the strength of this fluctuation attraction is larger in systems with larger salt concentration and a larger number of adsorption sites. 

\begin{figure*}[!t]
\subfloat[]{\includegraphics[width=0.33\textwidth]{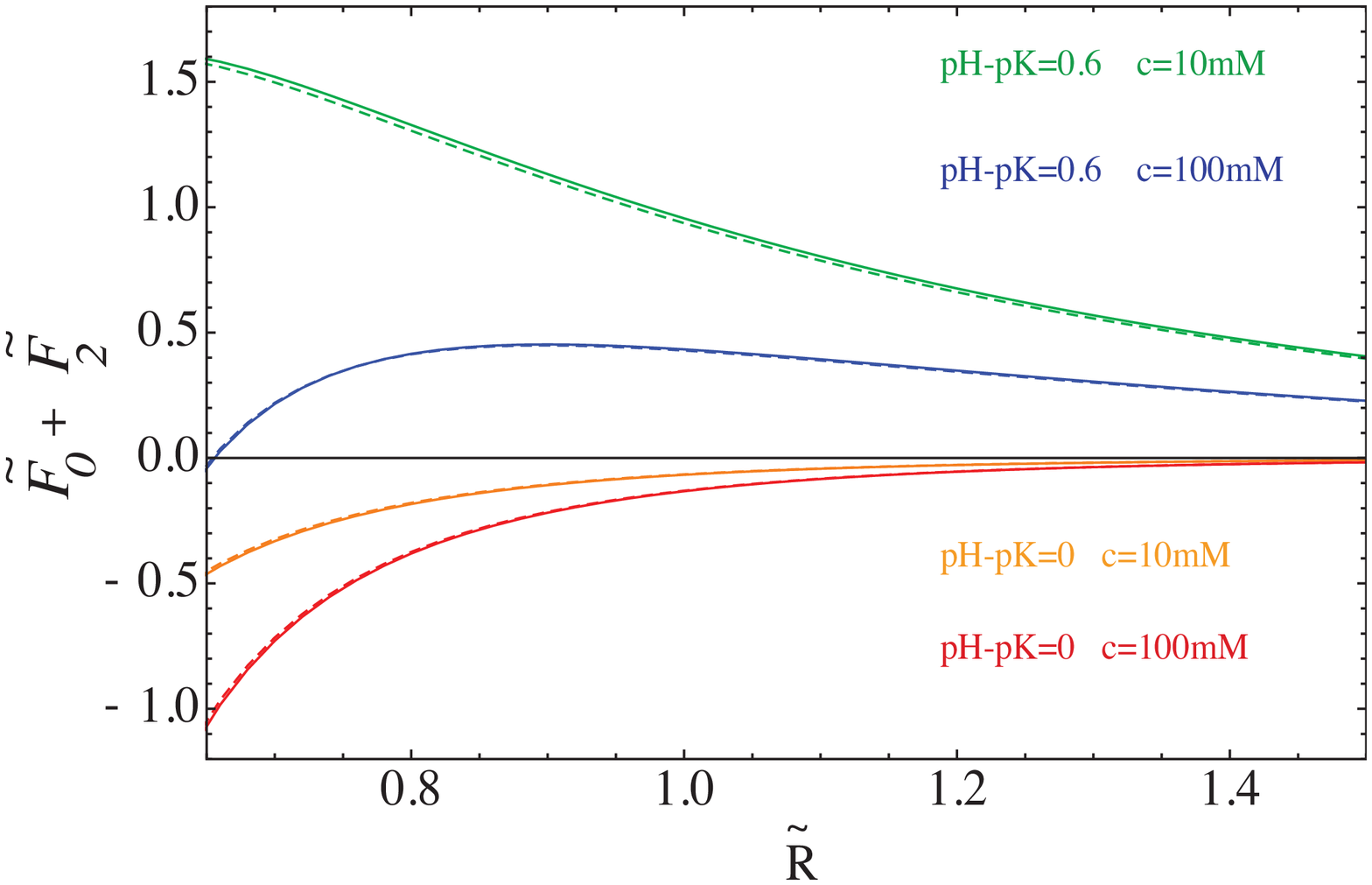}}\subfloat[]{\includegraphics[width=0.33\textwidth]{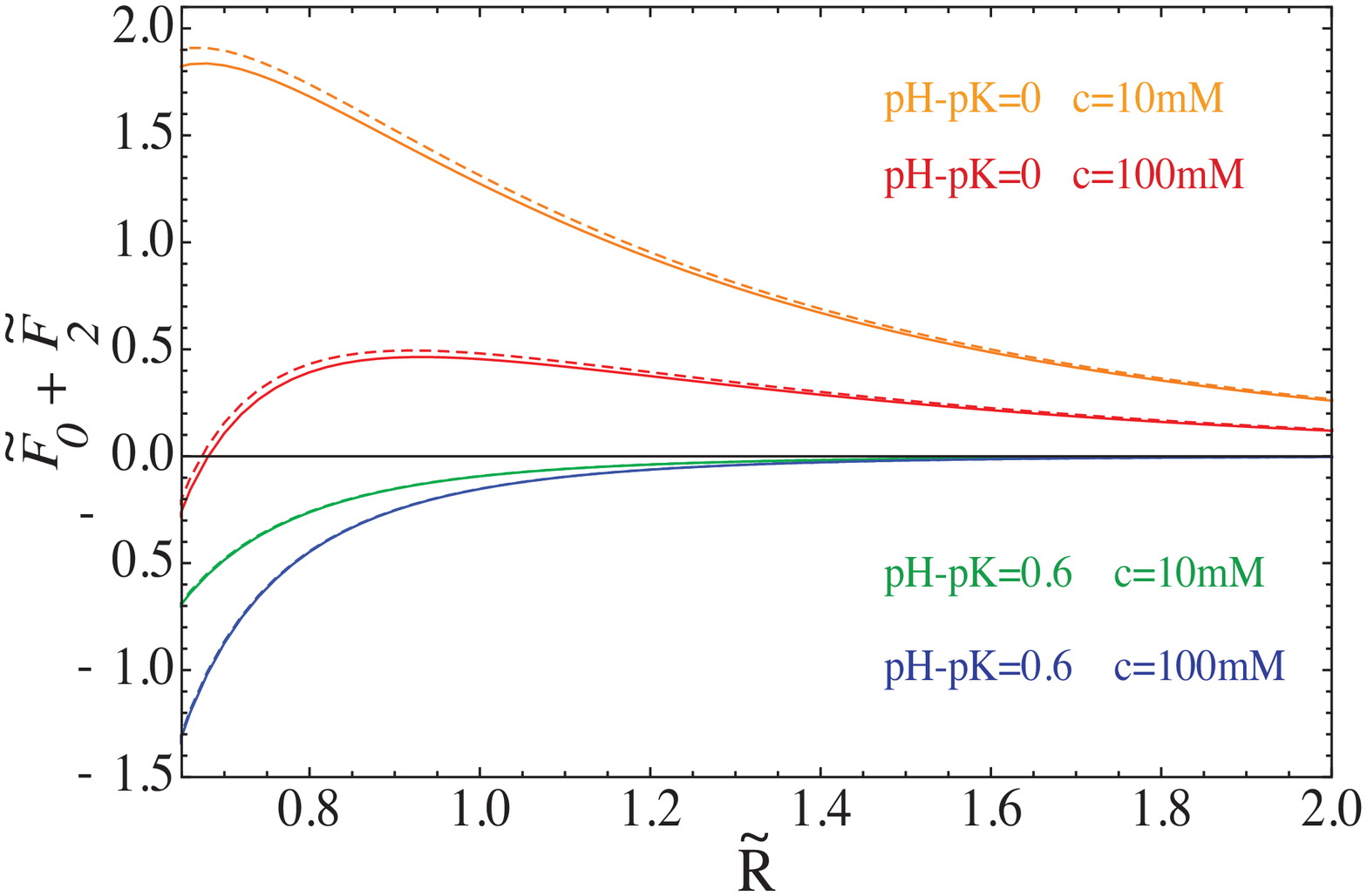}}\subfloat[]{\includegraphics[width=0.335\textwidth]{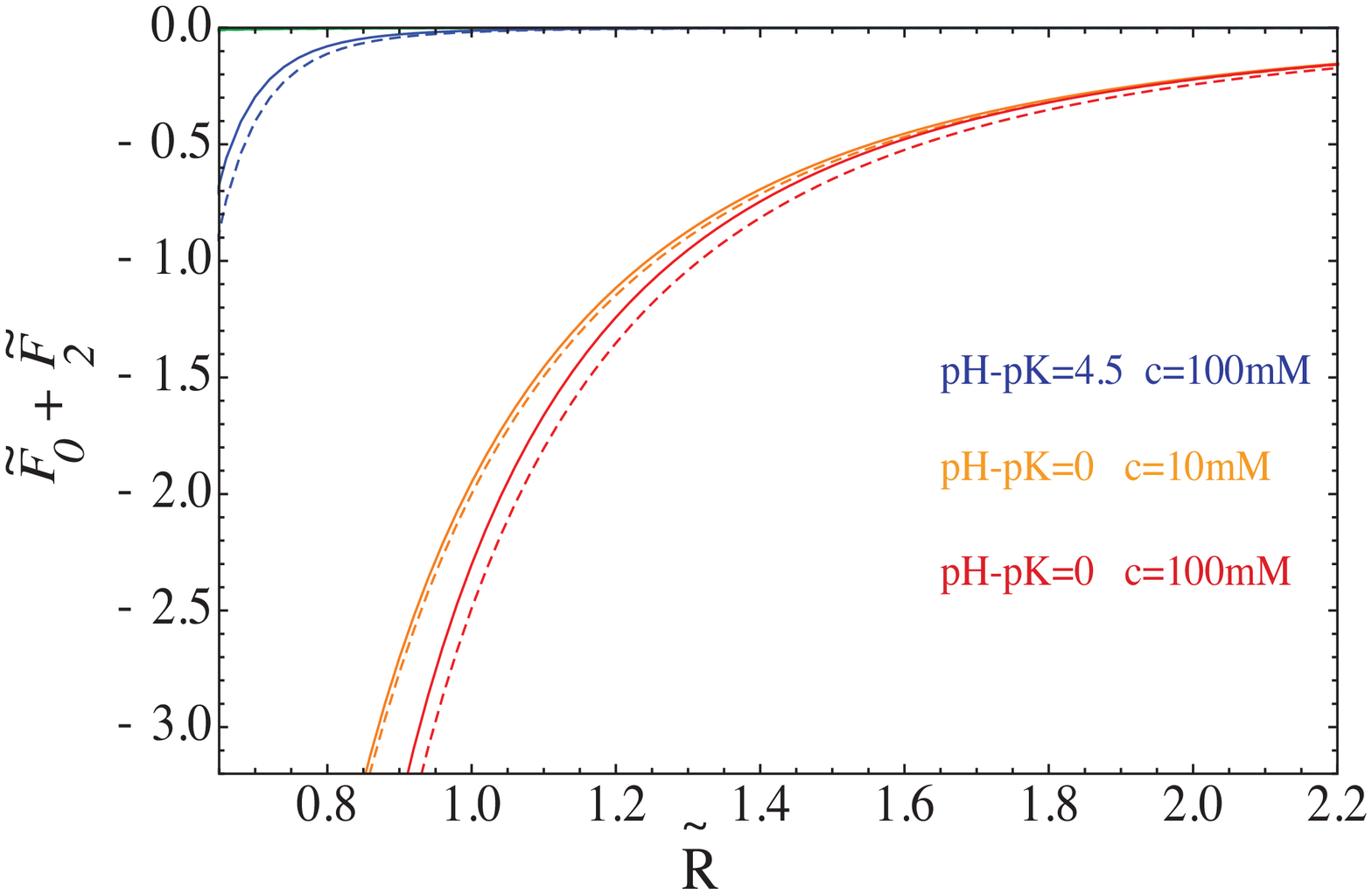}}
\caption{Total interaction force, obtained using the saddle-point approximation to evaluate the full partition function, (dashed lines) compared with numerical results, obtained using the exact evaluation of the full partition function, (solid lines). (a) fully symmetric system ($\alpha =2$); (b) semi-symmetric system with  $\alpha =5$; (c) asymmetric system. Each color corresponds to a different choice of parameters (number of adsorption sites $N$ and salt concentration $c$, and $pH-pK$) as indicated. The dimensionless diameter of the macroions is set to be $\tilde{a}=0.5$. }
\label{fig:fig5}
\end{figure*}

Concerning the semi-symmetric system of macroions with both charges  spanning the same asymmetric interval, Fig. \ref{fig:fig2} (dashed lines), one discernes similar behavior of all averages as in the fully symmetric system. However here, the PZC is no longer determined by $pH=pK$, but is shifted,  meaning that the concentration of the positive ions close to the macroion surfaces is different from the concentration of protons in the bulk. The auto-correlation function as a function $pH-pK$ is not centered anymore on the PZC, but the asymmetric fluctuations do again appear at the PZC, Fig. \ref{fig:fig2} (b), where one can observe net attraction between the macroions, Fig. \ref{fig:fig22} (dashed lines).

The behavior of the completely asymmetric system is shown in Fig \ref{fig:fig3}, Fig \ref{fig:fig33}. Here, away from PZC, the first macroion is positive, the second neutral, or the first can be neutral, while the second can be negatively charged, depending on the value of $pH-pK$. In the region $-3\lessapprox pH-pK\lessapprox 3$ both macroions carry nonzero charge of opposite sign, and at $pH=pK$, the system is electroneutral as a whole, i.e. the sum of average charges is equal to zero, Fig \ref{fig:fig3}(a). The charge cross correlation function is always negative, Fig \ref{fig:fig3}(b) and one can observe only attraction, Fig \ref{fig:fig3}(c).  The number of adsorption sites has the biggest influence on the intensity of interaction. 

The fluctuation effect shows an interesting twist in this system: the interaction force as a function of separation shows attraction also when one of the macroions is charged and the other reaching its point of zero charge, see Fig. \ref{fig:fig33}(a). The origin of that attraction comes from the mean charge-induced charge interaction, see Fig. \ref{fig:fig33}(b), where one can observe non-zero product  $<\tilde{e}_i>^2(<\tilde{e}_j - <\tilde{e}_j>>^2)$ of non-zero charge $<\tilde{e}_i>^2$ and autocorelation function of zero charge $<\tilde{e}_j>$.  As it is the case in the symmetric system, here also for the same dimensionless separation the attraction is significantly stronger in a solution with larger salt concentration.  

\section{Discussion}\label{sec:4}

In the previous section we showed results obtained numerically using the exact evaluation of the full partition function. The aim of this section is to see, whether one can proceed analytically in order to get a better intuition about the behavior of the attractive interaction arising between identical macroions with fluctuating charge, so that it can be compared with the original KS result for the attractive components as well as the DH result for the repulsive component, respectively. In order to do so, we will evaluate the partition function, Eq. \ref{eq:aspartfunction}, introducing two different approximations,  the {\sl saddle-point} approximation and the {\sl "Gaussian"} approximation, comparing the ensuing approximative results with the exact ones. The approximations refer to the evaluation of the partition function Eq. \ref{eq:path} and not to the evaluation of the field Green's function, $G(\varphi_1,\varphi_2)$, which is always assumed to be of the DH form. All the approximations detailed below thus refer to the evaluation of the charge regulation part of the partition function.

\subsection{Saddle-point approximation}

The saddle-point approximation consists of finding the dominant contribution to the partition function,  corresponding to the minimum of the field action, which is then expanded around the minimum to the second order in deviation. The saddle-point approximation is usually referred to also as the mean-field approximation, but we need to distinguish the mean-field in the treatment of the charge regulation free energy with the PB mean-field, which refers to the interaction part. The procedure is detailed in Appendix \ref{sec:apsp}, where we derive expressions for the saddle-point free energy, as well as the fluctuation induced free energy from the second-order correction Eq. \ref{eq:b8}. With respect to that decomposition, one can distinguish the saddle-point interaction force, $\tilde{F}_{0}$, and the fluctuation component of the interaction force, $\tilde{F}_{2}$, with magnitudes given as:
\begin{eqnarray}
&&\tilde{F}_0=k\frac{1+\tilde{R}}{\tilde{R}^2}\tilde{a}^2e^{2\tilde{a}-\tilde{R}}\frac{(\phi_1^*-\frac{\tilde{a}}{\tilde{R}}e^{\tilde{a}-\tilde{R}}\phi_2^*)(\phi_2^*-\frac{\tilde{a}}{\tilde{R}}e^{\tilde{a}-\tilde{R}}\phi_1^*)}{\left(1-(\frac{\tilde{a}}{\tilde{R}})^2e^{-2(\tilde{R}-\tilde{a})}\right)^2}\nonumber\\
~\label{eq:f0}
\end{eqnarray}
and: 
\begin{eqnarray}
&&\tilde{F}_2=-\frac{1+\tilde{R}}{\tilde{R}^3}\frac{\tilde{a}^2e^{-2(\tilde{R}-\tilde{a})}}{h_1(\phi _1^*) h_2(\phi _2^*)-\frac{\tilde{a}^2}{\tilde{R}^2}e^{-2(\tilde{R}-\tilde{a})}}\label{eq:f2}.
\end{eqnarray}
Here $k=\frac{4\pi \epsilon \epsilon_0}{\beta e_0^2 \kappa }$, while $h_1(\phi _1^*)$ and $h_2(\phi _2^*)$ are defined as:
\begin{eqnarray}
h_1(\phi _1^*)&=&1+\frac{k\tilde{a}}{\alpha b N}e^{\tilde{a}} e^{-\phi_1^*}(b+e^{\phi_1^*})^2;\nonumber\\
h_2(\phi _2^*)&=&1+\frac{k\tilde{a}}{\alpha b N}e^{\tilde{a}}  e^{-\phi_2^*}(b+e^{\phi_2^*})^2,
\end{eqnarray}
with $\phi _1^*$ and $\phi _2^*$ the solutions of the saddle-point equations,  Eq. \ref{eq:b3}, \ref{eq:b4}, given in the Appendix \ref{sec:apsp}. Since, they are obtained numerically, this method does not give us a transparent analytical solution for the free energy and interaction force.

The sum of the saddle-point interaction force, $\tilde{F}_0$, and the fluctuation force, $\tilde{F}_2$, for symmetric, semi-symmetric and asymmetric systems are plotted as functions of separation $\tilde{R}$ and compared with results obtained with exact evaluation of the full partition function, Fig. \ref{fig:fig5}. One can notice that there is an excellent agreement between both results obtained using these different methods. Saddle-point method decouples the total force into a saddle-point part and a fluctuation part, one being repulsive and the other attractive, respectively, except for the asymmetric system, where there is no repulsion whatsoever, Fig. \ref{fig:fig5}(c).  They can be differentiated based on the separation scaling of the interaction free energy. In the first case it decays exponentially with $\tilde R$, while in the second it decays exponentially with $2 \tilde R$. The repulsive force decreases as the system is approaching the PZC, where it is identically equal to zero. In this regime the fluctuation component to the interaction force becomes dominant one. 

The main and important difference between the interactions calculated exactly or on the saddle-point level, is that the attractive component of the interaction force in the latter case, does not depend on $pH$, but is however sensitive and increases with the salt concentration, Fig. \ref{fig:fig5} (a), (b). The full pH dependence of the interaction is thus not described properly by the saddle-point approximation.

\subsection{Gaussian approximation}

In this case the analytical evaluation of the  partition function Eq. \ref{eq:aspartfunction}, is based on a Gaussian  approximation for the binomial coefficient, and it is presented in Appendix \ref{sec:approx}. 

The partition function in this case also decouples into two separate contributions, of which one decays exponentially with $\tilde R$, and the other one decays exponentially with $2 \tilde R$. We will again refer to them as the "mean" and the "fluctuation" part of the interaction force, using the same notation as for the saddle-point approximation. One should note here that on this approximation level there is no real decoupling into the mean and fluctuation part. We differentiate them purely based on their separation scaling.

The mean interaction force, $\tilde{F}_{0}$, can be obtained as:
\begin{eqnarray}
&&\tilde{F}_{0}=k\frac{1+\tilde{R}}{\tilde{R}^2}\tilde{a}^2e^{2\tilde{a}-\tilde{R}}\frac{[(pH-pK)\ln{10}]^2}{\left(1+2\frac{k\tilde{a}}{ N}e^{\tilde{a}}+ \frac{\tilde{a}}{\tilde{R}}e^{-(\tilde{R}-\tilde{a})}\right)^2}\nonumber\\
~\label{eq:fcc}
\end{eqnarray}
and the fluctuation force, $\tilde{F}_{2}$, as:
\begin{eqnarray}
&&\tilde{F}_{2}=-\frac{1+\tilde{R}}{\tilde{R}^3}\frac{\tilde{a}^2 e^{-2(\tilde{R}-\tilde{a})}}{(1+\frac{4 k\tilde{a}}{\alpha N}e^{\tilde{a}} )^2-\frac{\tilde{a}^2}{\tilde{R}^2}e^{-2(\tilde{R}-\tilde{a})}}.\label{eq:fks}
\end{eqnarray}
Again both $\tilde{F}_{0}$ and $\tilde{F}_{2}$ are obtained in the same way and the separation into "mean" and "fluctuation" part is arbitrary. Nevertheless, the separation scaling of the two is the same as for the mean-field and fluctuation contribution in the case of the saddle-point approximation, making the nomenclature reasonable.

The general form of mean interaction force is given in Appendix \ref{sec:approx}, Eq. \ref{eq:fullfcc}, valid for all three systems considered: fully symmetric, semisymmetric and asymmetric. Because of its complexity, we display here only $\tilde{F}_{0}$ for the fully symmetric system, Eq. \ref{eq:fcc}. On the other side, the fluctuation force, Eq. \ref{eq:fks}, has the same, universal form for all three types of systems. One can compare these results,  Eq. \ref{eq:fcc}, Eq. \ref{eq:fks}, with those obtained using the saddle-point approximation, Eq. \ref{eq:f0}, Eq. \ref{eq:f2}. 

Clearly the fluctuation force in the Gaussian approximation corresponds exactly to the fluctuating force in the saddle-point approximation, if the saddle-point is taken at the PZC, $pH=pK$, and the mean-potentials are $\phi_1^*=\phi_2^*=0$. However, in general the two approximations do not coincide and thus we can not claim that $\tilde{F}_{2}$ is purely fluctuational in origin.

\begin{figure*}[!t]
\subfloat[]{\includegraphics[width=0.33\textwidth]{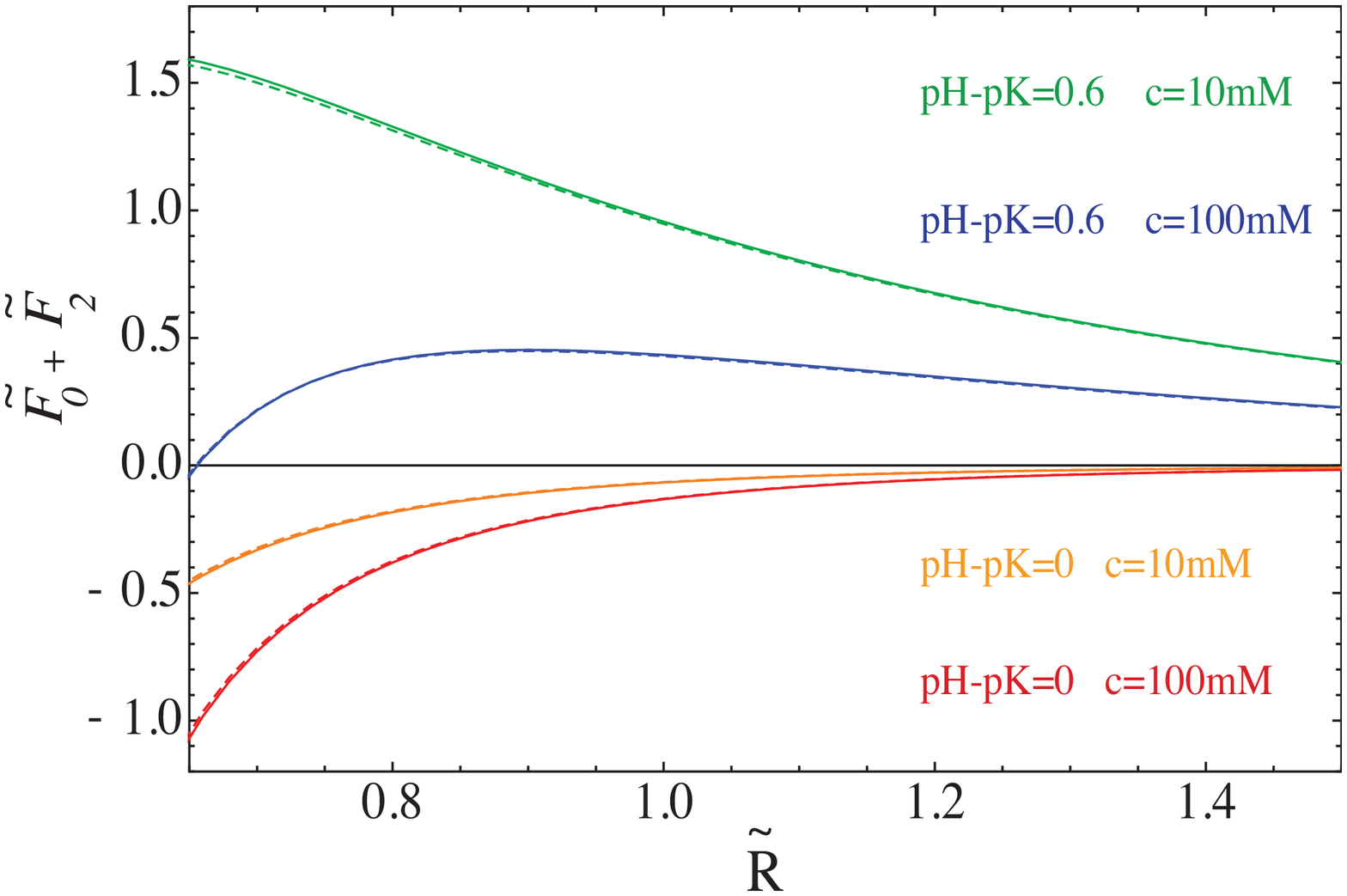}}\subfloat[]{\includegraphics[width=0.33\textwidth]{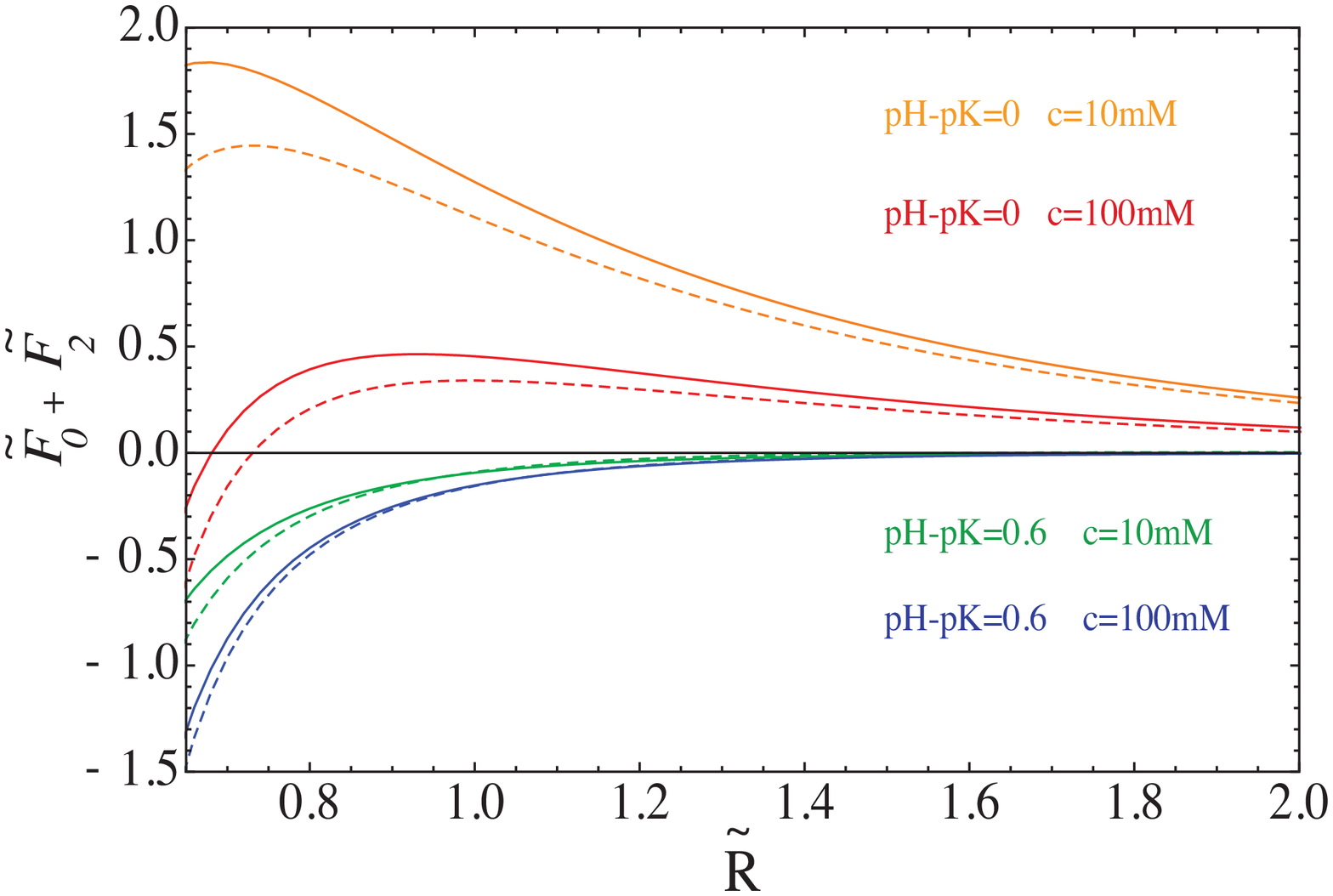}}\subfloat[]{\includegraphics[width=0.323\textwidth]{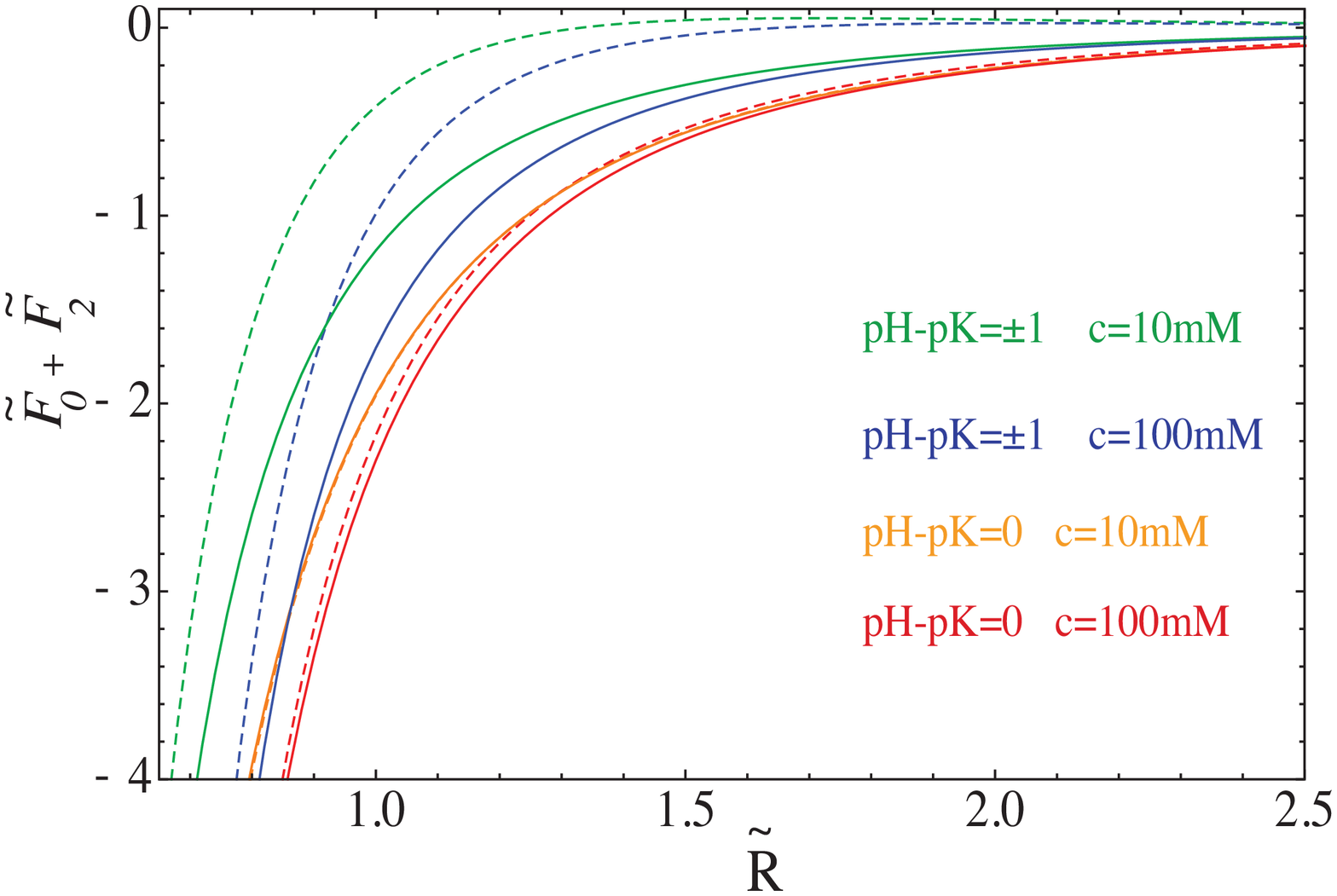}}\caption{Analytical results for the total force, obtained using approximative evaluation of full partition function, (dashed lines) are compared with numerical results, obtained using exact evaluation of full partition function, (solid lines). (a) fully symmetric system ($\alpha =2$); (b) semi-symmetric system with  $\alpha =5$; (c) asymmetric system. Each color suits to the corresponding choice of parameters (number of adsorption sites $N$ and salt concentration $c$, and $pH-pK$) as it is shown at figures. The dimensionless  diameter of the macroions is set to be $\tilde{a}=0.5$. }
\label{fig:fig4}
\end{figure*}

The mean and the fluctuation part  to the interaction force are plotted as functions of dimensionless separation $\tilde{R}$ in Fig. \ref{fig:fig4}. The total interaction force obtained in this way is compared with the one obtained using the exact evaluation of the partition function. For the fully symmetric system, the Gaussian approximation fits perfectly the exact results, Fig. \ref{fig:fig4}(a). A somewhat lesser agreement can be found in a semisymmetric system, Fig. \ref{fig:fig4}(b), while the analytical results do not work at all in the region away from PZC in the asymmetric system, Fig. \ref{fig:fig4}(c). 

In the fully symmetric system, the mean part of the interaction force is repulsive, decreasing on approach to the PZC, while in the asymmetric system, it is actually attractive as the macroions are on the average oppositely charged. On the other side, the fluctuation component to the interaction force is  attractive no matter what the symmetry of the system and the $pH$ of solution, while it does depend on the salt concentration.  Interestingly enough, on the Gaussian approximation level for the binomial coefficient the $pH$-dependence of the auto-correlation function again drops out completely, which is contrary to the full numerical evaluation of the charge auto-correlation function. 

\subsection{Comparison with DH and KS forms}

We now set our results agains the mean-field DH theory of interactions between point-like macroions, and against the KS theory of charge fluctuation forces.  Obviously without charge regulation the charge of both interacting macroions is fixed and the DH form of the interaction should be recovered. Setting $\alpha =0$ and $M=N$ in Eq. \ref{eq:aspartfunction}, one indeed get the DH interaction force between two well separated like-charged macroions in a salt solution:
\begin{equation}
\tilde{F} \approx \frac{N^2}{k} \frac{e^{-\tilde{R}}}{\tilde{R}}.\label{eq:fdh}
\end{equation}
Charge regulation, besides inducing attraction at the PZC, also introduces significant modifications in the mean-field interaction force, Eq. \ref{eq:fcc}, leading to its vanishing  at the PZC. In the limit of large separations, the charge-regulated interaction force  Eq. \ref{eq:fcc}, in fact scales as:
\begin{equation}
\tilde{F}_{0} \approx \frac{1}{\tilde{R}}k\tilde{a}^2e^{2\tilde{a}-\tilde{R}}\frac{[(pH-pK)\ln{10}]^2}{\left(1+2\frac{k\tilde{a}}{ N}e^{\tilde{a}}\right)^2},
\end{equation} 
clearly showing a strong dependence on the solution pH.
 
As for the fluctuation component of the interaction force for two spherical point-like macroions, we can cast its form in the  Gaussian approximation, going to a limit of large separation, Eq. \ref{eq:fks}, as  
\begin{equation}
\tilde{F}_{2}\approx -\frac{1}{\tilde{R}^2}\frac{\tilde{a}^2 e^{-2(\tilde{R}-\tilde{a})}}{(1+2\frac{k\tilde{a}}{N}e^{\tilde{a}} )^2}.\label{eq:fkss}
\end{equation} 
The charge auto-correlation function for the two macroions, $<\Delta \tilde{e}_1^2>=<(\tilde{e}_1-<\tilde{e}_1>)^2>$, is calculated analytically using the same Gaussian approximation and the following form is obtained:
\begin{equation}
<\Delta \tilde{e}_1^2><\Delta \tilde{e}_2^2>\approx \frac{k^2 \tilde{a}^2 e^{2\tilde{a}}}{(1+2\frac{k\tilde{a}}{N}e^{\tilde{a}})^2}.
\end{equation}
With this result the fluctuation component of the interaction force assumes the asymptotic form:
\begin{equation}
\tilde{F}_{2}\approx -\frac{e^{-2\tilde{R}}}{k^2\tilde{R}^2}<\Delta \tilde{e}_1^2><\Delta \tilde{e}_2^2>.\label{eq:fkkss}
\end{equation}  
This actually coincides exactly with the original Kirkwood-Schumaker result \cite{KS1, KS2} if we take into account the fact that they take the DH Green's function for two point charges with a finite size-scaling factor $e^{\tilde{a}}/(1+\tilde{a})$, so that we would have to multiply Eq. \ref{eq:fkkss} by $e^{-2\tilde{a}}(1+\tilde{a})^2$. Again we note that on this approximation level the $pH$-dependence of the auto-correlation function drops out completely, but is retained in the full numerical evaluation of the charge auto-correlation function.

\section{Protein-like macroions}\label{sec:5}

The general theory formulated above can be straightfrowardly applied to the interaction of protein-like macroions at large separations. In a protein, the amino acids (AAs) Asp, Glu, Tyr and Cys can be negatively charged, while Arg, Lys and His can carry a positive charge, all depending on the solution conditions. The respective pKs for the dissociation of the various amino acids are given at Table I \cite{Szleifer}.

In order to describe a protein macroion composed of these amino acids, one should write down the charge regulation free energy in the form:
\begin{eqnarray}
&&f_{p}(\varphi )=i\sum_j N_j M_j e_0 \varphi - kT \sum_j N_j M_j \ln{\left(1+b_j e^{i\beta e_0 \varphi }\right)} -\nonumber\\
&&- kT \sum_k N_k M_k \ln{\left(1+b_k e^{i\beta e_0 \varphi }\right)},
\end{eqnarray}
where $j$ stands for negative AAs  $j=\{Asp, Glu, Tyr, Cys\}$, while $k$ stands for positive ones $k=\{Arg, His, Lys\}$.  $N_j$, $N_k$ are the numbers of absorption sites on each positive and negative AAs and since each of these AAs has one adsorption site it will be set to $1$.  $M_j, M_k$ count how many times each of AAs occurs in the protein, and $b_j$, $b_k$ stand for $b_n=e^{-\ln{10}(pH-pK_n)}$, where $pK_n$ for each AA is given in Table I. For point-like macroions the spatial distribution of AAs on the surface of the protein is irrelevant and the above approximation is thus admissible.

\begin{figure*}[!t]
\subfloat[]{\includegraphics[width=0.34\textwidth]{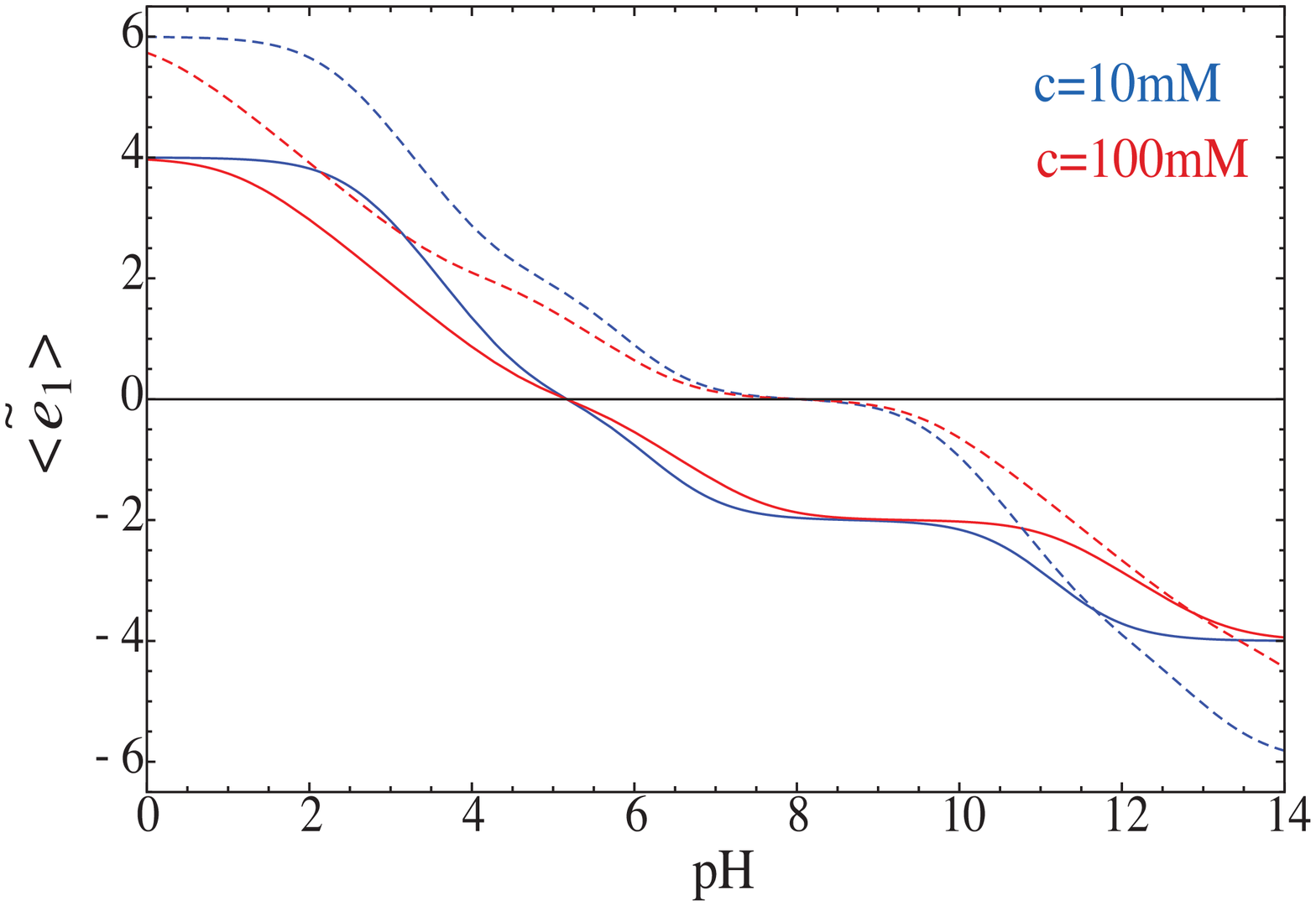}}\subfloat[]{\includegraphics[width=0.34\textwidth]{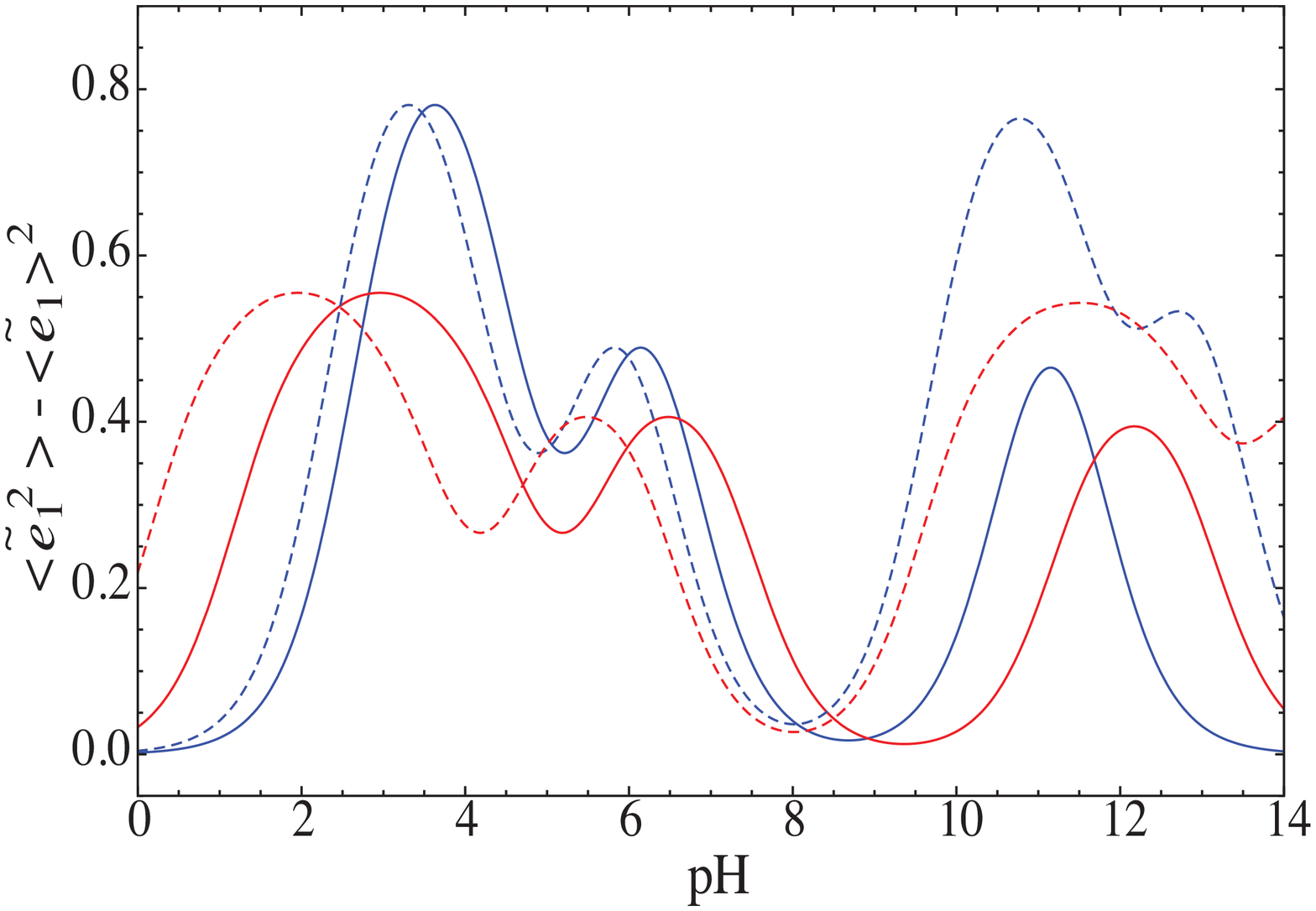}}
\subfloat[]{\includegraphics[width=0.34\textwidth]{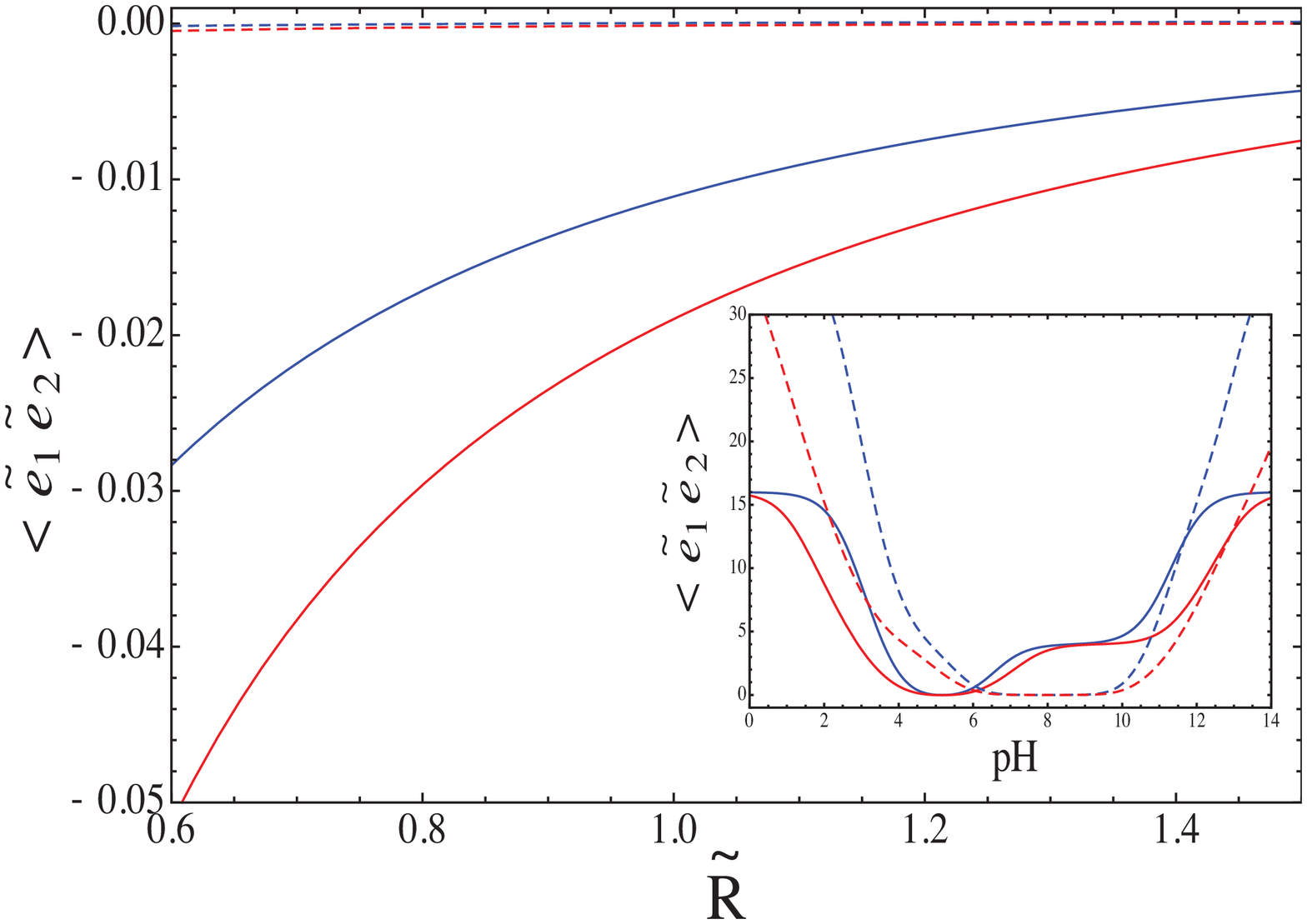}}
\caption{Generalized  system: (a) The average charge of macroions; (b) charge auto-correlation function; (c) charge-charge cross-correlation function. All results are obtained by using the exact numerical evaluation of the full partition function. Solid lines correspond to a system of two proteins, each of which consists of 2 Asp, 2 Glu, 2 Lys, 2 His, while dashed lines represent the system which has additional 2 Tyr and 2 Arg.  Blue color corresponds to the value of salt concentration $c=10mM$, while the red color corresponds to the $c=100mM$. The dimensionless diameter of the macorions is set to be $\tilde{a}=0.5$ and separation between them $\tilde{R}=1$. The functions bearing $\tilde{R}$-dependence are plotted at isoelectric point of two systems: $pH=5.15$ and $pH=7.87$ respectively.}
\label{fig:fig6}
\end{figure*}

\begin{figure}[!t]
\includegraphics[width=0.47\textwidth]{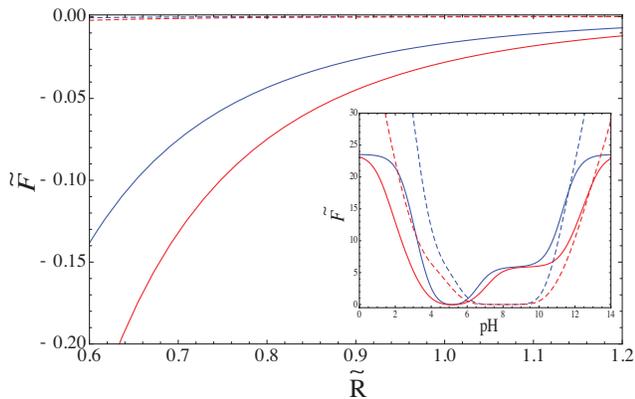}
\caption{Generalized  system:  interaction force. All results are obtained by using the exact numerical evaluation of the full partition function. Solid lines correspond to a system of two proteins, each of which consists of 2 Asp, 2 Glu, 2 Lys, 2 His, while dashed lines represent the system which has additional 2 Tyr and 2 Arg.  Blue color corresponds to the value of salt concentration $c=10mM$, while the red color corresponds to the $c=100mM$. The functions bearing $\tilde{R}$-dependence are plotted at isoelectric point of two systems: $pH=5.15$ and $pH=7.87$ respectively. The dimensionless diameter of the macroions is set to be $\tilde{a}=0.5$ and separation between them $\tilde{R}=1$}
\label{fig:fig66}
\end{figure}

\begin{table}
\begin{ruledtabular}
\begin{tabular}{cccccccc}
   & ASP & GLU & TYR & ARG & HIS & LYS & CYS \\
\hline
 pK & 3.71 & 4.15 & 10.10 & 12.10 & 6.04 & 10.67 & 8.14 \\
\end{tabular}
\caption{$pK$ values of amino-acids functional groups in dilute aqueous solution, after Ref.  \cite{Szleifer}.}
\end{ruledtabular}\label{tab:tab}
\end{table}

The partition function for the system composed of two protein-like macro-ions in a 1:1 salt solution, is derived in the same way as explained in Sec \ref{sec:2}, and is given in Appendix \ref{sec:ap2}. Since the evaluation of  Eq. \ref{eq:zprot}, is computationally time consuming, we consider only the behavior of two model systems, one (system I) composed of protein-like macro-ions consisting of  2 Asp, 2 Glu, 2 Lys, 2 His, and the other (system II) having 4 AAs more - 2 Tyr and 2 Arg. The results are shown in Fig. \ref{fig:fig6}. The protein charge, as a function of $pH$, spans a symmetric interval with constant  plateaus in the pH regions, that correspond to charging up an additional AA.  The cross-correlation function in general follows the pattern of plateaus of the average charge,  being positive  everywhere except at the PZC, where asymmetric charge distribution appears. The auto-correlation function and the charge cross-correlation show opposite signs, with one being positive and the other negative, respectively. 

Analyzing the behavior of the interaction force, one can see that two identical proteins mutually repel and that the strength of the interaction depends on pH in the solution, following closely the behavior of the charge cross-correlation function. The repulsion is smaller in a solution of higher salt concentration, since the salt screens the protein charge and reduces the interaction. The repulsion disappears at the PZC, where the attraction sets in, increasing with salt concentration at a fixed dimensionless separation between the proteins.  The attractive interaction is negligible for proteins composed of a larger number of amino-acids, which is not in correspondence with our previous results, where the attraction is larger for a larger number of adsorption sites. This can be explained by analyzing the average charge of the protein, Fig.  \ref{fig:fig6} (a), where one can observe a plateau of zero charge for the system II, which is not the case in system I, so it can be concluded that the strength of the fluctuation interaction depends on the rate of change of the charge of the macro-ion with pH, which of course depends on the type of the protein. 

This can be derived also formally by following Lund and J\" onsson \cite{Lund}. The fluctuation part of the interaction force, Eq. \ref{eq:fkkss},  is approximately proportional to the charge variance, which in its turn follows from the macroion capacitance $\cal C$, as
 \begin{equation}
<(\tilde{e} -<\tilde{e}>)^2> \sim {\cal C} = \frac{\partial \tilde e(\phi)}{ \partial (\beta e_0\phi)} = - \frac{1}{\ln{10} } \frac{\partial \tilde e(\phi)}{\partial pH},
 \end{equation}
as is clear also from Eq. \ref{vaeht}. The strength of the fluctuation interaction therefore depends on the rate of change of the mean charge of the macroion with pH, i.e. its capacitance. This can be clearly discerned from Fig \ref{fig:fig6}(b), where we observe that the system II has zero capacitance at its PZC, while system I has a non-zero capacitance at its PZC.  

\section{Conclusion}\label{sec:6}

We presented a theory describing electrostatic interaction between two spherical macroions, with non-constant, fluctuating charge, surrounded by a monovalent bathing salt solution.  The macro-ion charge fluctuations are described with the Parsegian-Ninham model of charge regulation, that effectively corresponds to a lattice gas surface dissociation free energy. Our theory is based on two approximations: one assumes the macroions as point-like, in the sense that the electrostatic potential on the surface of the macro-ion is uniform, and other treats the intervening salt solution on the Debye-H\" uckel level, assuming the electrostatic potential to be small, so that the Poisson-Boltzmann equation can be linearized. Choosing the proper charge regulation energy, we analyzed the behavior of three different systems that differ in the symmetry of charge distribution. These are: a symmetric system composed of two identical macroions with a symmetric as well as asymmetric charge regulation intervals, corresponding to the fully symmetric and semisymmetric cases, and an asymmetric system, composed of oppositely charged macroions, allowing the case of having one charged and one uncharged particle. 

We have shown that in charge regulated systems, asymmetrical charge fluctuations appear near the PZC, engendering strong attractive interactions of a general Kirkwood- Schumaker type, but with different functional dependencies as argued in their original derivation. The fluctuational nature of the Kirkwood- Schumaker interaction is consistent also with the fact that it arrises even between a charged and a charge neutral object, in the vicinity of the $pH$ where the charged macro-ion becomes neutral itself. This is the case studied also in the context of the PB theory within the constant charge regulation model, in fact corresponding to a linearized form of the full charge-regulation theory \cite{Borkovec-1,Borkovec-2}. In this limit too, the effects of charge regulation are crucial and lead to attraction. However, in the context of our approximations, the attractive interaction between a charged and a neutral surface stemms from the coupling between the net charge of one, and charge fluctuations of the other surface. Off hand one would tend to see the attraction in the constant charge regulation model as being grounded in the mean-field level but caution should be exercized here. In our case too the Green's function pertains to the DH mean-field level, and the attraction actually comes from the surface charge regulation. Constant charge regulation model must obviously capture some of the same physics. 

Bathing solution with its pH and ionic strength therefore plays an important role in charge regulated systems, and the interactions to which they are subject.  In all cases studied, the fluctuation attraction is larger for larger salt concentration in solution at the same dimensionless separation, while the repulsion is actually reduced at a fixed separation by increasing the salt concentration, consistent with the electrolyte screening effect.  Furthermore, a stronger attraction is found in systems composed of identical macroions having a larger number of of adsorption sites, giving rise to larger charge fluctuations. 

The theory, developed for toy models, was then applied to the case of protein-like macroions, with different dissociation constant for different chargeable amino acids. For protein electrostatic interactions their strength depends on the rate of change of the charge of the macro-ion with respect to the solution pH, i.e. the molecular capacitance of the macroion, which is protein specific and connected with the capacitance of the protein charge distribution. Apart from this, salt concentration enhances the attraction between protein-like macroions, as is evidenced also in simulations and experiments in the case of e.g. lysozyme in monovalent salt solutions \cite{ll,curtis}. In fact understanding the details of the protein-protein interaction is our main motivation for developing further our theoretical approach, specifically the relation between the KS interaction, the patchiness effects and van der Waals interactions between proteins in electrolyte solutions.

 \section{Acknowledgments}

N. A. acknowledges the financial support by the Slovenian Research Agency under the young researcher grant. R.P. acknowledges the financial support by the Slovenian Research Agency under the grant P1-0055. He would also like to thank Prof. Michal Borkovec and Dr. Gregor Trefalt for illuminating discussions on the subject of charge regulation and electrostatic interactions between charge-regulated macroions.

\section{Appendix}

\subsection{Path-integral formalism}\label{sec:app}

The field propagator at points $\ \vec{r}_1$ and  $\ \vec{r}_2$ is defined as:
\begin{eqnarray}
G(\varphi_1,\varphi_2)&=&\int_{\varphi(\vec{r}_1)=\varphi_1}^{\varphi(\vec{r}_2)=\varphi_2}{\cal D}[\varphi(\vec{r})]\delta(\varphi(\vec{r}_1)-\varphi_1)\delta(\varphi(\vec{r}_2)-\varphi_2)\nonumber\\
&\times &\exp{\Bigg[-\frac{1}{2}\int d\vec{r}d\vec{r}' \varphi(\vec{r})G^{-1}(\vec{r},\vec{r})\varphi(\vec{r}')\Bigg]}\
\end{eqnarray}
where $G^{-1}(\vec{r},\vec{r}')$ is the usual Debye-H\" uckel kernel of the form  \cite{RudiandCo}:
\begin{equation}
G^{-1}(\vec{r},\vec{r}') = -\varepsilon_0 \left( \nabla \varepsilon(\vec{r}) \nabla - \varepsilon(\vec{r}) \kappa^2\right) \delta (\vec{r} - \vec{r}'),
\end{equation}
where $\kappa $ is the inverse Debye length. Using the delta function in integral representation:
\begin{eqnarray}
\delta(\varphi(\vec{r}_1)-\varphi_1)&=&\int \frac{dk}{2\pi}e^{ik(\varphi(\vec{r}_1)-\varphi_1)}=\nonumber\\
&&\int \frac{dk}{2\pi} e^{-ik\varphi_1+ik\int d\vec{r}\rho_1(\vec{r})\varphi(\vec{r})}
\end{eqnarray}
where $\ \rho_1(\vec{r})=\delta(\vec{r}-\vec{r}_1)$, one can rewrite the propagator as:
\begin{eqnarray}
&&G(\varphi_1,\varphi_2)=\int dke^{-ik\varphi_1}\int dk'e^{-ik'\varphi_2}\int{\cal D}[\varphi(\vec{r})]\nonumber\\
&&\exp{\!\!\Bigg[-\frac{1}{2}\int d\vec{r}d\vec{r}' \varphi(\vec{r})G^{-1}(\vec{r},\vec{r})\varphi(\vec{r}')\!+\!i\!\!\int \!\!t(\vec{r})\varphi(\vec{r})d^3\vec{r}\Bigg]}\nonumber\\
~
\end{eqnarray}
where $t(\vec{r})$ stands for $t(\vec{r})=k\rho_1(\vec{r})+k'\rho_2(\vec{r})$. After integration over the field, one obtains:
\begin{eqnarray}
&&G(\varphi_1,\varphi_2)=\frac{1}{\det{G^{-1}(\vec{r},\vec{r'})}}\int dke^{-ik\varphi_1}\int dk'e^{-ik'\varphi_2}\nonumber\\
&&\exp{\left(-\frac{1}{2}\int d\vec{r}d\vec{r}' t(\vec{r})G(\vec{r},\vec{r'})t(\vec{r}')\right)}=\nonumber\\
&&=\frac{1}{\det{G^{-1}(\vec{r},\vec{r'})}}\int_{-\infty}^{+\infty}\int_{-\infty}^{+\infty}dkdk'e^{-ik\varphi_1-ik'\varphi_2}\nonumber\\
&&\times e^{-\frac{1}{2}k^2G(\vec{r}_1,\vec{r}_1)}\times e^{-\frac{1}{2}k'^2G(\vec{r}_2,\vec{r}_2)}\times e^{-kk'G(\vec{r}_1,\vec{r}_2)}\nonumber\\
~
\end{eqnarray}
If one introduces a 2D vector $(k,k')$, this integral can be rewritten as:
\begin{eqnarray}
G(\varphi_1,\varphi_2)&=&\frac{1}{\det{G^{-1}(\vec{r},\vec{r'})}}\int \int dk dk'e^{-i (\varphi_1, \varphi_2)\left(\begin{array}{c} k \\ k' \end{array}\right)}\nonumber\\
&&e^{-\frac{1}{2}\left(k,  k'\right)\Big(
\begin{matrix}
G(\vec{r}_1,\vec{r}_1) &  G(\vec{r}_1,\vec{r}_2)\\
G(\vec{r}_1,\vec{r}_2) &  G(\vec{r}_2,\vec{r}_2)
\end{matrix}\Big)\left(\begin{array}{c} k \\ k' \end{array}\right)}
\end{eqnarray}
Since this is a Gaussian integral, it can be evaluated explicitly:
\begin{eqnarray}
&&G(\varphi_1,\varphi_2)=\nonumber\\
&&\exp{\Big[-\frac{\beta }{2}\left(\varphi_1, \varphi_2\right)\Big(
\begin{matrix}
G(\vec{r}_1,\vec{r}_1) &  G(\vec{r}_1,\vec{r}_2)\\
G(\vec{r}_1,\vec{r}_2) &  G(\vec{r}_2,\vec{r}_2)
\end{matrix}\Big)^{-1}\left(\begin{array}{c} \varphi_1 \\ \varphi_2 \end{array}\right)\Big]}.\nonumber\\
~
\end{eqnarray}

\subsection{Saddle-point approximation}\label{sec:apsp}

The partition function Eq \ref{eq:path} can be evaluated using the saddle-point method, consisting of minimization of the field action $\frac{\partial A}{\partial \phi }=0$, where the action can be written in the form:
\begin{equation}
A(\phi_1,\phi_2)=f_1(\phi_1)+g(\phi_1,\phi_2)+f_2(\phi_2),
\end{equation}
with $g$ the logarithm of the Green's function, given as:
\begin{equation}
g(\phi_1,\phi_2)=-\frac{1}{2}\left(\phi_1, \phi_2\right)\Big(
\begin{matrix}
\tilde{G}(\vec{r}_1,\vec{r}_1) &  \tilde{G}(\vec{r}_1,\vec{r}_2)\\
\tilde{G}(\vec{r}_1,\vec{r}_2) &  \tilde{G}(\vec{r}_2,\vec{r}_2)
\end{matrix}\Big)^{-1}\left(\begin{array}{c} \phi_1 \\ \phi_2 \end{array}\right).
\end{equation}
The saddle-point equations are obtained as:
\begin{eqnarray}
&&N-\alpha N \frac{b e^{-\phi _{1}}}{1+be^{-\phi _{1}}}+\phi_{1} \frac{4\pi \epsilon \epsilon_0}{\beta e_0^2 \kappa }\frac{1}{\frac{e^{-2\tilde{a}}}{\tilde{a}^2}-\frac{e^{-2\tilde{R}}}{\tilde{R}^2}}\frac{e^{-\tilde{a}}}{\tilde{a}}-\nonumber\\
&&\phi_{2} \frac{4\pi \epsilon \epsilon_0}{\beta e_0^2 \kappa }\frac{1}{\frac{e^{-2\tilde{a}}}{\tilde{a}^2}-\frac{e^{-2\tilde{R}}}{\tilde{R}^2}}\frac{e^{-\tilde{R}}}{\tilde{R}}=0;\label{eq:b3}\\
&&M-\alpha N \frac{b e^{-\phi _{2}}}{1+be^{-\phi _{1/2}}}+\phi_{2} \frac{4\pi \epsilon \epsilon_0}{\beta e_0^2 \kappa }\frac{1}{\frac{e^{-2\tilde{a}}}{\tilde{a}^2}-\frac{e^{-2\tilde{R}}}{\tilde{R}^2}}\frac{e^{-\tilde{a}}}{\tilde{a}}-\nonumber\\
&&\phi_{1} \frac{4\pi \epsilon \epsilon_0}{\beta e_0^2 \kappa }\frac{1}{\frac{e^{-2\tilde{a}}}{\tilde{a}^2}-\frac{e^{-2\tilde{R}}}{\tilde{R}^2}}\frac{e^{-\tilde{R}}}{\tilde{R}}=0.\label{eq:b4}
\end{eqnarray}
Solutions of these  equations are denoted as $\phi_1^*$ and $\phi_2^*$. If one sets $M=0$, $\alpha =1$, one  deals with an asymmetric system, for  $M=N$, $\alpha =2$, one deals with a fully symmetric system,  while the choice  $M=N$, $\alpha >2$ defines a symmetric system with an asymmetric interval of fluctuating charge, i.e. a semi-symmetric system.

The action can be expanded around the SP solution up to the second order in deviation from $\phi_1^*$ and $\phi_2^*$, yielding:
\begin{eqnarray}
&&A(\phi_1,\phi_2)=f_1(\phi_1^*)+g(\phi_1^*, \phi_2^*)+f_2(\phi_2^*)+\nonumber\\
&&\frac{1}{2}\frac{\partial^2A(\phi_1, \phi_2)}{\partial \phi_1^2}\vert_{\phi_1^*, \phi_2^*}\delta \phi_1^2+\frac{\partial^2A(\phi_1, \phi_2)}{\partial \phi_1\partial \phi_2}\vert_{\phi_1^*, \phi_2^*}\delta \phi_1\delta \phi_2+\nonumber\\
&&\frac{1}{2}\frac{\partial^2A(\phi_1, \phi_2)}{\partial \phi_2^2}\vert_{\phi_1^*, \phi_2^*}\delta \phi_2^2,
\end{eqnarray}
where $f_{1/2}(\phi _{1/2}^*)$ are given as:
\begin{equation}
f_{1/2}(\phi_{1/2}^*)=- M\phi_{1/2}^* -\alpha N\ln{(1+b e^{-\phi_{1/2}^*})}
\end{equation}
If we denote second derivatives in the equation above as $A_{11}$, $A_{12}$ and $A_{22}$ respectively, we will have:
\begin{eqnarray}
A_{11}&=&-\alpha  N b \frac{e^{- \phi_1^*}}{(1+b e^{- \phi_1^*})^2}-\frac{4\pi \epsilon \epsilon_0}{\beta e_0^2 \kappa }\frac{1}{\frac{e^{-2\tilde{a}}}{\tilde{a}^2}-\frac{e^{-2\tilde{R}}}{\tilde{R}^2}} \frac{e^{-\tilde{a}}}{\tilde{a}}\nonumber\\
A_{22}&=&-\alpha  N b \frac{e^{- \phi_2^*}}{(1+b e^{- \phi_2^*})^2}-\frac{4\pi \epsilon \epsilon_0}{\beta e_0^2 \kappa }\frac{1}{\frac{e^{-2\tilde{a}}}{\tilde{a}^2}-\frac{e^{-2\tilde{R}}}{\tilde{R}^2}} \frac{e^{-\tilde{a}}}{\tilde{a}}\nonumber\\
A_{12}&=&\frac{4\pi \epsilon \epsilon_0}{\beta e_0^2 \kappa }\frac{1}{\frac{e^{-2\tilde{a}}}{\tilde{a}^2}-\frac{e^{-2\tilde{R}}}{\tilde{R}^2}} \frac{e^{-\tilde{R}}}{\tilde{R}},
\end{eqnarray}
so that the saddle-point and the fluctuation free energy are equal to:
\begin{eqnarray}
&& \beta {\cal{F}}_0=- [f_1(\phi_1^*)+g(\phi_1^*, \phi_2^*)+f_2(\phi_2^*)]
\end{eqnarray}
and
\begin{eqnarray}
 &&\beta {\cal{F}}_2=-\ln{\frac{\det{A_0}}{\det{A}}}\label{eq:b8}
\end{eqnarray}
 where $A_0$ is a matrix,  related to the partition function of the unperturbed system, with the elements:
 \begin{eqnarray}
A_{11}^0=\frac{\partial^2A_0(\phi_1, \phi_2)}{\partial \phi_1^2}&=&-\frac{4\pi \epsilon \epsilon_0}{\beta e_0^2 \kappa }\frac{1}{\frac{e^{-2\tilde{a}}}{\tilde{a}^2}-\frac{e^{-2\tilde{R}}}{\tilde{R}^2}} \frac{e^{-\tilde{a}}}{\tilde{a}};\nonumber\\
A_{22}^0=\frac{\partial^2A_0(\phi_1, \phi_2)}{\partial \phi_2^2}&=&-\frac{4\pi \epsilon \epsilon_0}{\beta e_0^2 \kappa }\frac{1}{\frac{e^{-2\tilde{a}}}{\tilde{a}^2}-\frac{e^{-2\tilde{R}}}{\tilde{R}^2}} \frac{e^{-\tilde{a}}}{\tilde{a}};\nonumber\\
A_{12}^0=\frac{\partial^2A_0(\phi_1, \phi_2)}{\partial \phi_1\partial \phi_2}&=&\frac{4\pi \epsilon \epsilon_0}{\beta e_0^2 \kappa }\frac{1}{\frac{e^{-2\tilde{a}}}{\tilde{a}^2}-\frac{e^{-2\tilde{R}}}{\tilde{R}^2}} \frac{e^{-\tilde{R}}}{\tilde{R}}.
\end{eqnarray}

Finally, the saddle-point interaction force and the force due to the fluctuations around the saddle-point are given as:
\begin{eqnarray}
\tilde{F}_0&=&\frac{4\pi \epsilon \epsilon_0}{\beta e_0^2 \kappa }\frac{1+\tilde{R}}{\tilde{R}^2}\tilde{a}^2e^{2\tilde{a}-\tilde{R}}\times\nonumber\\
&&\frac{(\phi_1^*-\frac{\tilde{a}}{\tilde{R}}e^{\tilde{a}-\tilde{R}}\phi_2^*)(\phi_2^*-\frac{\tilde{a}}{\tilde{R}}e^{\tilde{a}-\tilde{R}}\phi_1^*)}{\left(1-(\frac{\tilde{a}}{\tilde{R}})^2e^{-2(\tilde{R}-\tilde{a})}\right)^2}\\
\tilde{F}_2&=&-\frac{1+\tilde{R}}{\tilde{R}^3}\frac{\tilde{a}^2e^{-2(\tilde{R}-\tilde{a})}}{h_1(\phi _1^*) h_2(\phi _2^*)-\frac{\tilde{a}^2}{\tilde{R}^2}e^{-2(\tilde{R}-\tilde{a})}}
\end{eqnarray}
where:
\begin{eqnarray}
h_1(\phi _1^*)&=&1+\frac{4\pi \epsilon \epsilon_0\tilde{a}}{\beta e_0^2 \kappa N \alpha b}e^{\tilde{a}} e^{-\phi_1^*}(b+e^{\phi_1^*})^2\nonumber\\
h_2(\phi _2^*)&=&1+\frac{4\pi \epsilon \epsilon_0\tilde{a}}{\beta e_0^2 \kappa N\alpha b}e^{\tilde{a}}  e^{-\phi_2^*}(b+e^{\phi_2^*})^2
\end{eqnarray}
The saddle-point and the fluctuation force are plotted as functions of dimensionless separation $\tilde{R}$ in the Fig. \ref{fig:fig5}.

\subsection{Gaussian approximation}\label{sec:approx}

The partition function Eq. \ref{eq:aspartfunction} can be evaluated analytically, if one takes a Gaussian approximation for the binomial coefficient:
\begin{equation}
\left(\begin{array}{c} \alpha N \\ n \end{array}\right)=\frac{2^{\alpha N}}{\sqrt{\frac{\pi \alpha  N}{2}}}e^{-\frac{(\alpha N-2n)^2}{2\alpha N}}.\label{eq:napx0}
\end{equation}
After substitution $x=\alpha N-2n$ and $x'=\alpha M-2n'$, summation can be transformed into the integral, when one assumes $N \gg 1$, so that the partition function becomes: 
\begin{eqnarray}
{\cal{Z}}&=&\int _{-\infty}^{\infty} dx\int _{-\infty}^{\infty} dx' e^{\frac{1}{2}(x+x')(pH-pK)\ln{10}}\nonumber\\
&&e^{-\frac{1}{2\alpha N}(x^2+x'^2)}e^{-\beta {\cal{F}}(x, x',\tilde{R})},
\end{eqnarray}
where
\begin{eqnarray}
&&{\cal{F}}(x,x',\tilde{R})=\frac{e_0^2\kappa }{8\pi \epsilon\epsilon_0}\times \nonumber\\
&&\Big[\frac{e^{-\kappa a}}{a}\Big( (x+N(2-\alpha ))^2+(x'+M(2-\alpha ))^2\Big)+\nonumber\\
&&2\frac{e^{-\kappa R}}{R}(x+N(2-\alpha ))(x'+M(2-\alpha ))\Big].
\end{eqnarray}
This is a general Gaussian-type integral and can be calculated analytically, but since the solution is too cumbersome, it is not displayed here. The interaction force then follows as a sum of the mean contribution to the force and the fluctuation force as:
\begin{widetext}
\begin{eqnarray}
&&\tilde{F}_{0}=k\tilde{a}^2e^{2\tilde{a}-\tilde{R}}\frac{1+\tilde{R}}{\tilde{R}^2}\frac{[(pH-pK)\ln{10}]^2}{\left(1+2\frac{k\tilde{a}}{ N}e^{\tilde{a}}+ \frac{\tilde{a}}{\tilde{R}}e^{-(\tilde{R}-\tilde{a})}\right)^2}+\frac{(\alpha -2)k\tilde{a}^2e^{2\tilde{a}-\tilde{R}}}{\alpha^2N(1+\frac{4k\tilde{a}e^{\tilde{a}}}{\alpha N})^2}\frac{1+\tilde{R}}{\tilde{R}^2}\Bigg(-\frac{2\alpha (N+M)(pH-pK)\ln{10}}{\left(1+\frac{1}{(1+\frac{4k\tilde{a}e^{\tilde{a}}}{\alpha N})^2}\frac{\tilde{a}}{\tilde{R}}e^{-(\tilde{R}-\tilde{a})}\right)^2}+\nonumber\\
&&\frac{4\alpha M(\alpha N\!-\!1)(pH\!-\!pK)\ln{10}\frac{1}{1+\frac{4k \tilde{a}e^{\tilde{a}}}{\alpha N}}e^{-(\tilde{R}-\tilde{a})}}{\left(1-\frac{1}{(1+\frac{4k\tilde{a}e^{\tilde{a}}}{\alpha N})^2}\frac{\tilde{a}^2}{\tilde{R}^2}e^{-2(\tilde{R}-\tilde{a})}\right)^2}+\frac{ (4\alpha M\!-\!8) (1+\frac{1}{(1+\frac{4k\tilde{a}e^{\tilde{a}}}{\alpha N})^2}\frac{\tilde{a}^2}{\tilde{R}^2}e^{-2(\tilde{R}-\tilde{a})})\!-\!(\alpha \!-\!2)(1+\alpha \frac{M^2}{N^2})\frac{4N}{1+\frac{4k\tilde{a}e^{\tilde{a}}}{\alpha N}}\frac{\tilde{a}}{\tilde{R}}e^{-(\tilde{R}-\tilde{a})}}{\left(1-\frac{1}{(1+\frac{4k\tilde{a}e^{\tilde{a}}}{\alpha N})^2}\frac{\tilde{a}^2}{\tilde{R}^2}e^{-2(\tilde{R}-\tilde{a})}\right)^2}\Bigg);\nonumber\\
&&\tilde{F}_{2}=-\frac{1+\tilde{R}}{\tilde{R}^3}\frac{\tilde{a}^2 e^{-2(\tilde{R}-\tilde{a})}}{(1+\frac{4}{\alpha}\frac{4\pi \epsilon \epsilon_0\tilde{a}}{\beta e_0^2 \kappa N}e^{\tilde{a}} )^2-\frac{\tilde{a}^2}{\tilde{R}^2}e^{-2(\tilde{R}-\tilde{a})}}.\nonumber\\
~\label{eq:fullfcc}
\end{eqnarray}
\end{widetext}
The mean contribution to the force  and fluctuation force are plotted as a  functions of separation $\tilde{R}$ and results are presented at the Fig. \ref{fig:fig4}. We note that the nomenclature "mean" and "fluctuation" do not have the same meaning in the context of the Gaussian approximation as they do in the saddle-point approximation. In fact in the former the interaction free energy can not be consistently separated into a mean and fluctuation types. We use this separation based on the dimensionless separation scaling.

\subsection{Protein-like macroions}\label{sec:ap2}

The partition function for the system of two point-like proteins immersed in  monovalent salt solution and containing seven types of dissociable AAs, negatively charged  $\{ Asp, Glu, Tyr, Cys \}$ and positively charged $\{ Arg, His, Lys \}$, can be written as:
\begin{widetext}
\begin{eqnarray}
&&{\cal{Z}}= \prod_{\ell = 1,7} \sum_{i_{\ell}, i'_{\ell}}^{M^{\ell}_i, M^{\ell}_{i'}} b_{\ell}^{i_{\ell}+i'_{\ell}} \frac{ M^{\ell}_{i} !  }{ i_{\ell}!  (M^{\ell}_i - i_{\ell})!} \frac{M^{\ell}_{i'} !  }{ i'_{\ell}!  (M^{\ell}_{i'} - i'_{\ell})!} e^{-\beta {\cal F}_{pp}} .\nonumber\\
~
\end{eqnarray}
where ${\ell}$ runs through $\{ Asp, Glu, Tyr, Cys \}$ and $\{ Arg, His, Lys \}$, with:
\begin{eqnarray}
&&{\cal F}_{pp}=\frac{ e_0^2\kappa }{8\pi \epsilon\epsilon_0}  \Big[\frac{e^{-\tilde{a}}}{\tilde{a}} \left( \sum_{m}(M^{m}_i - i)^2+\sum_{m}(M^{m}_{i'} - i')^2\right)+ 2\frac{ e^{-\tilde{R}} }{\tilde{R}} \sum_{m}(M^{m}_i - i) \sum_{m}(M^{m}_{i'} - i')\Big],\label{eq:zprot}
\end{eqnarray}
\end{widetext}
where the unprimed/primed notations referred to the two protein macroions. 
$M^{\ell}_i$ counts how many times each of these seven amino-acids occurs in a protein, while $M^{m}_i$ is restricted on counting only negative amino acids. $b_{\ell}$ refer to the chemical energy of dissociation: $b_{\ell}=e^{-\ln{10}(pH-pK_{\ell})}$, where the intrinsic $pK_{\ell}$ for the seven dissociable amino-acids are given in the Table I. 

\bibliography{your-bib-file}
\section{REFFERENCES:}

\end{document}